\documentclass[journal]{IEEEtran}
\usepackage{cite}
\usepackage{amsmath,amssymb,amsfonts}
\usepackage{graphicx}
\usepackage{textcomp}
\usepackage{multirow}
\usepackage{tikz}

\usepackage{ctable}
\usepackage{cleveref}
\usepackage{amsmath,amssymb}
\usepackage{subcaption}
\usepackage[table]{xcolor}
\usepackage{url} 
\usepackage{pifont}
\usepackage{makecell}
\usepackage[final]{changes}
\makeatletter
\newcommand{\shortto}{\mathrel{\mathpalette\shortto@\relax}}
\newcommand{\shortto@}[2]{%
  \vcenter{\hbox{\scalebox{0.6}{$\m@th#1\to$}}}%
}
\makeatother

\def\journalname{}

\newcommand{\mcc}[1]{\multicolumn{1}{c}{#1}} % short for multicolumn-centeredhttps://www.overleaf.com/project/685197461a527915424ac7b3#

\newcommand*{\myrulefill}[3][]{%
  \makebox[#2]{#1%
    \leaders\hrule height \dimexpr.5ex+.2pt\relax depth \dimexpr -.5ex+.2pt\relax \hfill% Left rule
    \enskip{#3}\enskip% Text (and surrounding spaces)
    \leaders\hrule height \dimexpr.5ex+.2pt\relax depth \dimexpr -.5ex+.2pt\relax \hfill\kern0pt}% Right rule
}

\def\BibTeX{{\rm B\kern-.05em{\sc i\kern-.025em b}\kern-.08em
    T\kern-.1667em\lower.7ex\hbox{E}\kern-.125emX}}
\begin{document}
\bstctlcite{BSTcontrol}

\title{Deep Learning Strain Estimation: Is Physics-Based Simulation the Solution?}

\author{Thierry~Judge, Nicolas~Duchateau, Andreas~Østvik, Khuram~Faraz, Anders~Austlid~Taskén, Sigve~Karlsen, Thor~Edvardsen, Harald~Brunvand, Md~Abulkalam~Azad, Havard~Dalen, Bjørnar~Grenne, Gabriel~Kiss, Pierre-Yves Courand, Lasse~Lovstakken, Pierre-Marc~Jodoin, and~Olivier~Bernard %\vspace{-0.6cm}
\thanks{T. Judge (thierry.judge@usherbrooke.ca) and P.-M. Jodoin, Dept. of Computer Science, University of Sherbrooke, Sherbrooke,  Canada.}
\thanks{T. Judge, K. Faraz, N. Duchateau, P.-Y. Courand, and O. Bernard, INSA, Université Lyon 1, CNRS UMR 5220, Inserm U1206, CREATIS, Villeurbanne, France.}
\thanks{N. Duchateau and O. Bernard, Institut Universitaire de France (IUF).}
% \thanks{A.-A. Taskén, G. Kiss, A. Østvik, and  L. Løvstakken are with Norwegian University of Science and Technology, Trondheim, Norway}
% \thanks{L. Løvstakken is with St Olavs University Hospital, Trondheim, Norway}
\thanks{P.-Y. Courand, Cardiology Dept., Hôpital Croix-Rousse, Hospices Civils de Lyon, Lyon, France, and the Cardiology Dept., Hôpital Lyon Sud, Hospices Civils de Lyon, Lyon, France.}
\thanks{A.-A. Taskén, and G. Kiss, Dept. of Computer Science, Faculty of Information Technology and Electrical Engineering, Norwegian University of Science and Technology (NTNU), Trondheim, Norway}
\thanks{A. Østvik, Md A. Azad, H. Dalen, B. Grenne, and L. Lovstakken, Dept. of Circulation and Medical Imaging, NTNU, Trondheim, Norway}
% \thanks{L. Løvstakken, NTNU, St Olavs Hospital, Trondheim, Norway}
% \thanks{A. Østvik, Md A. Azad, H. Dalen, and B. Grenne, NTNU, Department of Circulation and Medical Imaging, Trondheim, Norway}
\thanks{S. Karlsen, and H. Brunvand, Department of Medicine, Hospital of Southern Norway, Arendal, Norway}
% \thanks{Md A. Azad, H. Dalen, and B. Grenne, Clinic of Cardiology and Cardiothoracic Surgery, St. Olavs Hospital, Trondheim, Norway}
\thanks{A. Østvik, Md A. Azad, H. Dalen, and B. Grenne, Dept. of Cardiology and Cardiothoracic Surgery, St. Olavs Hospital, Trondheim, Norway}
\thanks{A. Østvik, Dept. of Health Research, SINTEF Digital, Trondheim, Norway}
\thanks{H. Dalen, Dept. of Medicine, Levanger Hospital, Nord-Trøndelag Hospital Trust, Levanger, Norway}
\thanks{T. Edvardsen, Dept. of Cardiology, Oslo University Hospital, Rikshospitalet and the Faaculty of Medicine, University of Oslo, Norway.}
\thanks{This work was supported in part by the NSERC’s Discovery Grants and Canada Graduate Scholarships-Doctoral programs, the FRQNT Doctoral Training Scholarships, the French National Research Agency (LABEX PRIMES [ANR-11-LABX-0063] of Université de Lyon, and ORCHID [ANR-22-CE45-0029-01] projects) and by the Fédération Française de Cardiologie - FOCACHIA project 2025-2026.} %For the purpose of open access, the authors have applied a CC BY public copyright license to any Author Accepted Manuscript (AAM) version arising from this submission.}
}

\maketitle

%%%%%%%%%%%%%%%%%%%%%%%%%%%%%%%%%%%%%%%%%%%%%%%%%%% Abstract%%%%%%%%%%%%%%%%%%%%%%%%%%%%%%%%%%%%%%%%%%%%%%%%%%%%%%%%%%%%%%%%%%%%%

\begin{abstract}
Speckle tracking echocardiography (STE) is the clinical standard for myocardial strain estimation. Despite good performance on global strain (GLS), its accuracy for regional strain remains limited, even though this biomarker is highly relevant for early diagnosis and the characterization of subtle abnormalities.
%potential for early diagnosis. 
% Deep learning has been studied as an alternative to this classical approach, but it is limited by the difficulty in obtaining reliable motion references. Two approaches have been proposed: learning from STE-derived labels on real images and using physics-based simulation to generate synthetic videos with known motion. However, existing simulators rely on simplified models of time-varying speckle and cardiac motion, producing sequences with limited realism that deviate from clinical data. 
Deep learning is a promising alternative, but its development is constrained by the lack of reliable motion references. Existing solutions rely either on STE-derived labels or on simulations generated by physics-based models, but these synthetic sequences still have limited realism compared with clinical data.
In this paper, we propose a novel 
simulation strategy %simulator 
that incorporates speckle decorrelation measures from real videos and uses an iterative refinement process to improve the motion realism in the simulations. We created an open-source photorealistic dataset of 1,478 videos with reference motion, which was used to train an echocardiographic motion estimation algorithm. The proposed method achieves unmatched performance on global and regional strain, notably reaching a GLS variability of 1.42\% in an inter-expert setting compared to 1.78\% for the clinical reference.
~\footnote{Code and database will be made public upon acceptance of this paper.}
\end{abstract}


%%%%%%%%%%%%%%%%%%%%%%%%%%%%%%%%%%%%%%%%%%%%%%%%%%% INTRODUCTION%%%%%%%%%%%%%%%%%%%%%%%%%%%%%%%%%%%%%%%%%%%%%%%%%%%%%%%%%%%%%%%%%%%%%
 % \vspace{-0.8cm}
\section{Introduction}
\label{sec:introduction}

Echocardiography is among the most widely recommended imaging modalities 
to quantify %for the quantification of 
cardiac structure and function in routine clinical practice~\cite{langRecommendationsCardiacChamber2015}. Segmentation of the cardiac chambers and myocardial wall enables the extraction of key volumetric and mass measurements. 
Tracking %In addition, tracking 
the myocardial wall across image sequences allows the estimation of local myocardial motion and myocardial deformation (also referred to as \emph{strain}), a sensitive marker of early dysfunction~\cite{cikesEjectionFractionIntegrative2016}. Nonetheless, while segmentation has reached clinical expert-level performance through deep learning~\cite{olaisen-ehj-ci-2023}, myocardial tracking 
%of the left ventricle (LV) remains an open challenge.
remains an open challenge even for the left ventricle (LV).

The current clinical standard, speckle tracking echocardiography (STE), relies on standard block matching and optical flow estimation without the use of 
deep learning. %deep learning–based techniques. 
However, it assumes temporal coherence of speckle patterns~\cite{smisethMyocardialStrainImaging2025a}, 
which %an assumption that 
is only valid over a limited number of consecutive frames within the cardiac cycle. In practice, this assumption is frequently violated due to speckle decorrelation induced by noise, imaging artifacts, fast cardiac motion, and out-of-plane motion. This leads to tracking errors that must be compensated for through extensive manufacturer-dependent tuning and regularization. As a result, myocardial strain estimation remains unreliable, particularly at the regional and local levels, despite its high relevance for assessing subtle disease patterns that are not accessible through global (i.e., myocardium-averaged) strain~\cite{mihosSpeckleTrackingStrainEchocardiography2025}.

Deep learning offers a potential path to overcome these limitations. Two major strategies have been investigated in the literature so far: training motion estimation networks directly on STE-derived labels from real images~\cite{azadEchoTrackerAdvancingMyocardial2024,chernyshovLowComplexityPoint2025}, or using simulated ultrasound data as ground truth~\cite{evainMotionEstimationDeep2022,taskenEstimationSegmentalLongitudinal2025}. \replaced{The first approach may inherit some of the biases of STE, which could limit its ability to outperform the method it aims to replace.}{The first approach inherently inherits the biases of STE and therefore cannot outperform the method it aims to replace.}
The second approach is constrained by the realism of existing simulation pipelines, which assume temporally consistent speckle patterns and derive motion from simplified heart models~\cite{burmanLargescaleSimulationRealistic2024} or 2D+t segmentations~\cite{evainMotionEstimationDeep2022}. These simplifications lead to unrealistic textures and limited motion variability in the generated synthetic images, resulting in a substantial domain gap with respect to the targeted real clinical data.

In this work, we go one step further by developing a simulation pipeline %designed 
to generate ultrasound videos with high spatial and temporal realism, specifically 
tailored for training deep learning–based methods. Our main contributions are: %summarized as follows:
\begin{itemize}
    \item A consolidated simulation strategy that explicitly models speckle decorrelation and generates synthetic ultrasound sequences closer to those from
    routine clinical imaging.
    \item An iterative dataset refinement strategy that leverages real patient data to progressively improve simulation realism and to generate a large-scale training dataset with physiologically plausible cardiac motion.
    
    % \item An adaptation of CoTracker~\cite{karaevCoTrackerItBetter2025}, a recent high-performing motion estimation method, to enable the tracking of arbitrary speckle patterns in echocardiography. The resulting method, hereafter referred to as \textbf{TAS-Net} for {\em Track Any Speckle Network}, is specifically tailored to the challenges of echocardiographic data.

    \item A new motion estimation model {\em Track Any Speckle Network} in the myocardium (\textbf{TAS-Net}), based on CoTracker~\cite{karaevCoTrackerItBetter2025} and specifically tailored to the challenges of echocardiographic data, to enable the tracking of arbitrary speckle patterns.

    % \item A comprehensive generalization study demonstrating that training on the proposed realistic synthetic dataset enables robust transfer to real echocardiographic videos and across different acquisition views.
    
\end{itemize}

Using 1,478 real echocardiographic videos spanning a wide range of textures and motion patterns, we generate a fully synthetic dataset with reference motion, which will be made openly available to support open and reproducible science. We demonstrate that this dataset enables the training of a high-performing motion estimation algorithm. Through extensive validation of both global and regional strain across multiple cohorts of real patients, we show that our method outperforms %performs \replaced{better than}{on par with, or better than,} 
conventional STE and recent deep learning–based approaches, while avoiding their respective limitations. To the best of our knowledge, this work is the first to achieve such performance using exclusively simulated data, thereby enabling a new generation of deep learning–based methods for global and regional strain estimation in echocardiography.

\begin{figure*}[tp]
\centering
\includegraphics[width=0.9\linewidth]{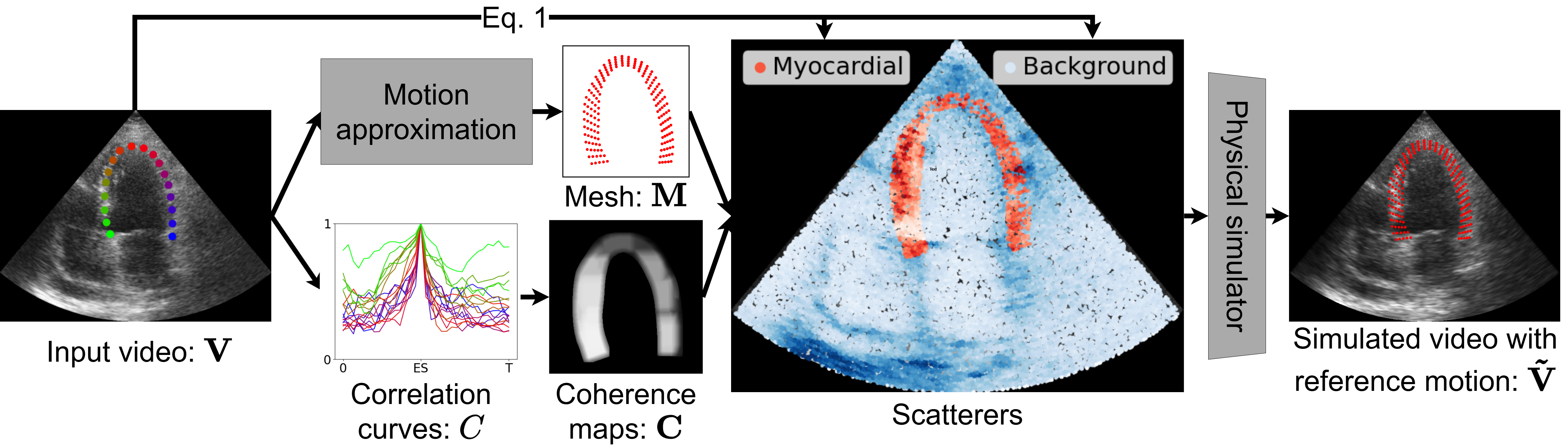}
% \caption{Overview of the proposed simulation pipeline. Starting from an original echocardiographic video and a subset of mesh centerline points, speckle correlation is estimated at each point to generate a dynamic coherence map. Cardiac motion is approximated during the iterative refinement process using either a 2D+t segmentation-based approach or a motion estimation algorithm. The resulting mesh and coherence maps control the generation of background and myocardial scatterers, which are used by the physical simulator to produce the synthetic images together with the corresponding reference motion.}
\caption{Overview of the proposed simulation pipeline. Starting from an echocardiographic video and mesh approximating the myocardium motion, speckle correlation is estimated at each point (only a subset is depicted) to generate a dynamic coherence map. The resulting mesh and coherence maps control the generation of background and myocardial scatterers, which are used by the physical simulator to produce the synthetic images.}
\label{fig:pipeline}
\end{figure*}

\section{Related Work}
\label{sec:related}

%\subsection{Motion estimation}
% \subsection{Motion estimatiofrom image sequences}
% \subsection{Motion estimation in natural images}

In recent years, deep learning has surpassed classical computer vision methods for optical flow estimation in natural videos. Key limitations, such as the scarcity of labeled training data, have been addressed through the use of large-scale synthetic datasets. Initial models such as PWC-Net~\cite{sunPwcnetCnnsOptical2018} and RAFT~\cite{teedRAFTRecurrentAllPairs2020} predicted dense optical flow between pairs of frames. Recently, research has shifted from dense frame-to-frame optical flow to sparse point-based tracking to better handle occlusions and long-range temporal dependencies, motivating the Track Any Point (TAP) task~\cite{doerschTapvidBenchmarkTracking2022}.
% for which multiple models have been proposed~\cite{harleyParticleVideoRevisited2022,zhengPointodysseyLargescaleSynthetic2023,doerschTapirTrackingAny2023}.
Being heavily driven by large-scale synthetic training datasets~\cite{greffKubricScalableDataset2022,zhengPointodysseyLargescaleSynthetic2023}, 
several %many 
models have been proposed~\cite{harleyParticleVideoRevisited2022,zhengPointodysseyLargescaleSynthetic2023,doerschTapirTrackingAny2023}.

% From a methodological perspective, Dosovitskiy \textit{et al.} first introduced the FlyingChairs dataset together with a fully convolutional neural network for optical flow estimation~\cite{dosovitskiyFlownetLearningOptical2015}. Subsequently, PWC-Net reintroduced principles from classical computer vision, including pyramidal processing, warping, and cost volume computation, leading to improved performance~\cite{sunPwcnetCnnsOptical2018}. RAFT further advanced the state of the art by leveraging 4D correlation volumes and iteratively refining the predicted optical flow~\cite{teedRAFTRecurrentAllPairs2020}. More recently, research has shifted from dense frame-to-frame optical flow to sparse point-based tracking to better handle occlusions and long-range temporal dependencies, motivating the Track Any Point (TAP) task introduced with the TAP-Vid benchmark and a baseline model~\cite{doerschTapvidBenchmarkTracking2022}. Since then, multiple models have been proposed, advancing point tracking performance~\cite{harleyParticleVideoRevisited2022,zhengPointodysseyLargescaleSynthetic2023,doerschTapirTrackingAny2023}.

Among recent advances, CoTracker stands out as one of the best-performing point-tracking models~\cite{karaevCoTrackerItBetter2025,karaevCotracker3SimplerBetter2025}. It is a transformer-based architecture that performs joint tracking of multiple points. After extracting spatial features using a convolutional neural network, CoTracker encodes a token for each query point and applies cross-track and cross-time attention mechanisms to iteratively refine trajectory estimates, achieving state-of-the-art tracking accuracy. 
% Progress in point tracking has also been closely linked to the availability of large-scale and increasingly realistic synthetic training datasets~\cite{greffKubricScalableDataset2022,zhengPointodysseyLargescaleSynthetic2023}. 
% Beyond purely synthetic data, recent approaches have further incorporated real videos through self-supervised learning~\cite{doerschBootstapBootstrappedTraining2024} and pseudo-labeling strategies~\cite{karaevCotracker3SimplerBetter2025}.

% \subsection{Myocardial motion estimation}
% \label{sec:myo_motion}

%While classical optical flow and point-tracking methods have largely been superseded by deep learning approaches in natural image processing, this transition has not yet occurred for %left ventricular myocardial motion estimation in cardiac imaging,including for the left ventricle, which is the most investigated cardiac chamber.
In clinical practice, speckle tracking echocardiography primarily relies on block-matching algorithms to estimate apparent myocardial motion~\cite{garciaIntroductionSpeckleTracking2018}. To mitigate limitations inherent to real echocardiographic image sequences, most notably speckle decorrelation, commercial tools rely on extensive manufacturer-specific tuning and regularization to produce smooth and physiologically plausible motion fields. This reliance constitutes a major limitation in terms of computational transparency and result reproducibility across vendors, an issue that has been a long-standing concern~\cite{intervendor}.

In recent years, deep learning has been increasingly adopted to provide data-driven solutions for myocardial motion estimation. As in natural image analysis, a major challenge in training deep learning models lies in the absence of reliable ground-truth motion. Several methods have explored learning from STE-derived labels. EchoTracker~\cite{azadEchoTrackerAdvancingMyocardial2024} employed a coarse-to-fine iterative tracking framework inspired by TAPIR~\cite{doerschTapirTrackingAny2023} to estimate point trajectories along the myocardial centerline. Similarly, 
\mbox{MyoTracker}~\cite{chernyshovLowComplexityPoint2025} proposed a simplified variant of CoTracker, removing the sliding window and iterative refinement mechanisms to enable fast and efficient tracking of right ventricular wall motion. While effective, such approaches can inherit the biases and limitations of STE-based supervision.

An alternative strategy relies on synthetic ultrasound data to provide motion ground truth. Evain \textit{et al.}~\cite{evainMotionEstimationDeep2022} proposed an adaptation of PWC-Net trained on 98 synthetic echocardiographic sequences derived from the CAMUS dataset and generated using the SIMUS simulator~\cite{garciaSIMUSOpensourceSimulator2022}, demonstrating improved performance compared to block-matching–based baselines. More recently, Taskén \textit{et al.}~\cite{taskenEstimationSegmentalLongitudinal2025} compared RAFT and CoTracker for myocardial motion estimation in transesophageal echocardiography using a similar synthetic framework. %, augmented with a simple controlled levels of speckle decorrelation. 
Their results showed superior performance of CoTracker over RAFT.

Beyond network architectures, recent efforts have focused on improving the realism and scale of synthetic echocardiographic datasets. Among them, %recent advances, 
Burman \textit{et al.}~\cite{burmanLargescaleSimulationRealistic2024} introduced a large-scale dataset 
of %comprising 
1,296 synthetic echocardiographic sequences generated from 50 patients from the CAMUS dataset using an advanced biophysical heart model (CircAdapt)~\cite{artsAdaptationMechanicalLoad2005} coupled with the COLE ultrasound simulator~\cite{gaoFastConvolutionbasedMethodology2009}. 
% This work highlights the increasing importance of realistic simulation pipelines for advancing deep learning–based myocardial motion estimation.

\section{Method}

Our work focuses on estimating 
LV %left ventricular 
myocardial motion from 2D echocardiographic image sequences, specifically apical 2-, 3-, and 4-chamber views. From the estimated motion, myocardial strain can then be derived with established clinical standards~\cite{brady22}. 
Our objective %The objective of this work 
is to track a set of points representing myocardial motion throughout a video $\mathbf{V} \in [0,1]^{T \times H \times W}$, composed of $T$ frames of height $H$ and width $W$. To capture the full extent of myocardial motion while leveraging recent advances in joint point tracking, we represent the myocardium using a structured mesh \mbox{$M \in \mathbb{R}^{l \times r \times 2}$}, with $l$ longitudinal and $r$ radial points. This mesh is initialized at a given frame and subsequently propagated over time, yielding a spatio-temporal mesh \mbox{$\mathbf{M} \in \mathbb{R}^{T \times l \times r \times 2}$}. 

To avoid the biases and limitations of STE-based methods, we propose in this section a new physics-based simulation pipeline specifically designed to generate synthetic echocardiogram videos with highly realistic temporal speckle patterns. %Inspired by the iterative training process used in the well-established SAM model for image segmentation, %  
It %Our pipeline 
relies on a fully automated refinement strategy to generate large-scale synthetic datasets with efficient reference myocardial motion. Building on this unique synthetic dataset, we propose a point-tracking model adapted from the CoTracker architecture, with several critical modifications designed to achieve high performance on real echocardiographic sequences.

\subsection{A new strategy for modeling speckle decorrelation}
\label{sec:speckle_decorrelation_strategies}

Our simulation pipeline relies on three complementary strategies designed to generate ultrasound textures with increasingly realistic time-varying speckle. First, we build upon a well-established simulation framework introduced in~\cite{evainMotionEstimationDeep2022} (referred to as Strategy 1). We then incorporate explicit speckle decorrelation mechanisms to model controlled temporal variations of speckle patterns (Strategy 2). Finally, using an effective two-step refinement procedure, we further adjust the simulated time-varying speckle such that the resulting speckle decorrelation matches the one observed in a real echocardiographic sequence used as template (Strategy 3).

%The objective of the simulation pipeline is to generate synthetic videos endowed with reference motion locally, to serve as labels for training the tracking algorithms. These simulations rely on real ultrasound sequences together with approximations of the underlying physical motion. While the resulting  synthetic  sequence remains visually similar to the input video, its speckle patterns evolve according to the ground-truth motion prescribed by the pipeline. We begin by presenting the  basic principles for this, inspired by the strategy introduced by Evain et al. \cite{evainMotionEstimationDeep2022}.

\subsubsection{Strategy 1 (S1): Base simulation pipeline}

%Although this pipeline has been previously introduced in the literature, we briefly summarize it here, as Strategies 2 and 3 build upon this baseline. T
As shown in \Cref{fig:pipeline}, our pipeline takes as input a real echocardiographic video $\mathbf{V}$, referred to as a template, and a temporal mesh $\mathbf{M}$ computed from the segmented myocardial region. This mesh provides a coarse approximation of the myocardial motion in the template, while its intrinsic regularization properties cause it to deviate from the true underlying motion. 

At the core of the proposed workflow is the SIMUS physical simulator~\cite{garciaSIMUSOpensourceSimulator2022}, specifically designed for ultrasound signal and image simulation. For each frame at time $t$ of the simulated sequence, the simulator takes as input a collection of acoustic point scatterers indexed by $i$, characterized by their spatial coordinates $(x_t^i,z_t^i)$ and their backscatter coefficients $\text{BSC}_t^i$. This coefficient is extracted from the corresponding real echocardiographic video according to:
\begin{equation}
\text{BSC}_t^i = (\mathbf{V}_{t, x_t^i, z_t^i})^\gamma \cdot \epsilon, \qquad \epsilon \sim \mathcal{N}(0,1),
\label{eq:bsc}
\end{equation}
where $\gamma$ is the gamma compression exponent and $\epsilon$ is a standard normal random variable.

As illustrated in ~\Cref{fig:pipeline} and following~\cite{evainMotionEstimationDeep2022}, two categories of scatterers are considered: myocardial scatterers, which move according to the temporal mesh $\mathbf{M}$, and background scatterers, which populate the remainder of the image. To distinguish between these two populations, a static coherence map $\mathbf{C}^s \in [0,1]^{T \times H \times W}$ is constructed by weighting the myocardial masks with a probability parameter $p \in [0,1]$. To ensure a smooth transition between the myocardium and the background, the coherence value is linearly decreased from the myocardial boundary outward. Using the end-systolic (ES) frame as %a 
reference—chosen because the myocardium is largely visible within the imaging sector—scatterers are uniformly distributed across the ultrasound sector at a density of five per square wavelength. Myocardial scatterers are randomly sampled according to the coherence map evaluated at their spatial locations. Their backscatter coefficients remain fixed over time, while their positions evolve according to their relative coordinates within the temporal mesh. In contrast, background scatterers remain spatially fixed but %have 
their backscatter coefficients 
are 
updated at each frame according to Eq.~(\ref{eq:bsc}). 
%In addition, 
A subset of background scatterers, selected based on the coherence map, is 
also 
randomly introduced within the 
myocardium %myocardial region 
throughout the cardiac cycle 
to enhance speckle realism.

Given the combined set of myocardial and background scatterers at each frame, together with the probe configuration, SIMUS~\cite{garciaSIMUSOpensourceSimulator2022} generates synthetic raw radio-frequency (RF) echoes. These signals are subsequently demodulated to obtain the complex baseband signal, commonly referred to as the in-phase/quadrature (I/Q) signal. The I/Q signals are then beamformed using a conventional delay-and-sum approach to reconstruct B-mode images~\cite{perrot2021}.

\subsubsection{Strategy 2 (S2): Integrating speckle decorrelation}

The simulation pipeline described above maintains a fixed ratio of coherent to incoherent scatterers, which in practice leads to \replaced{overly}{excessively} coherent speckle patterns throughout the simulated sequence. To overcome this limitation, we introduce the approach illustrated in ~\Cref{fig:pipeline}, which adaptively updates the myocardial-to-background scatterer ratio based on information extracted from the real video. This process requires quantifying the speckle correlation present in the input data. To this end, the correlation between the end-systolic (ES) frame and all other time instants of the sequence is computed for each mesh point. A $25 \times 25$ window centered at the mesh point in the ES frame serves as a reference, and normalized two-dimensional correlation is computed with a window centered at the corresponding point location in each frame $t$. The maximum correlation value is retained to account for possible mesh misalignments, yielding a correlation array $C \in \mathbb{R}^{T \times l \times r}$. 

These correlation curves are used to construct a dynamic coherence map, $\mathbf{C}^d \in [0,1]^{T \times H \times W}$, obtained by bilinear interpolation over the myocardial mask, as illustrated in ~\Cref{fig:pipeline}. This map governs the relative amplitude contributions of myocardial and background scatterers. To avoid flickering artifacts caused by the temporal appearance or disappearance of scatterers, both populations are maintained concurrently within the myocardial region. Although this locally doubles the scatterer density, the total BSC amplitude is preserved. \replaced{For each frame $t$, the BSC values of the myocardial $m$ and background $b$ scatterers are updated as: }{For each frame, the myocardial $m$ and background $b$ BSC values at each time $t$ are updated as: }

\begin{equation}
\left\{
\begin{aligned}
\mathrm{BSC}_{m,t}^i &=
\mathrm{BSC}_{m,ES}^i \cdot \mathbf{C}^d_{t, x_t^i, z_t^i}, \\[4pt]
\mathrm{BSC}_{b,t}^i &=
\mathrm{BSC}_{b,t}^i \cdot \left(1 - \mathbf{C}^d_{t, x_t^i, z_t^i}\right).
\end{aligned}
\right.
\end{equation}

%\begin{eqnarray}
%    \text{BSC}_{m,t}^i&=&\text{BSC}_{m,ES}^i \cdot \mathbf{C}^d_{t, x^i_t,z_t^i} \quad 
%    \\
%    \text{BSC}_{b,t}^i &=&\text{BSC}_{b,t}^i \cdot (1 - \mathbf{C}^d_{t, x^i_t,z_t^i}).
%\end{eqnarray}
%
This modulation of myocardial and background contributions induces time-varying speckle decorrelation that better matches real ultrasound behavior.

\subsubsection{Strategy 3 (S3): Two-step refinement procedure}

While the strategy described above improves the realism of the synthetic sequences, discrepancies in speckle decorrelation remain when compared with the original template sequences. To further align the simulated speckle decorrelation with that observed in the templates, we introduce a two-step refinement.
%Despite these improvements, limitations remain in this enhanced pipeline.
%the enhanced pipeline exhibits remaining limitations. 
% Factors beyond the balance of coherent and incoherent scatterers—including variations in intensity between ES and other instants—may artificially increase speckle correlation in regions where decorrelation should occur. For this reason, we introduce a refinement stepas follows.
First, a synthetic sequence is generated using Strategy 2, and the corresponding correlation curves, $C_{sim}$, are computed from the simulated output. The pointwise discrepancy between the target and simulated correlations is then defined as $C - C_{sim}$. A corrected set of input correlation curves is obtained according to $\tilde{C} = C + a \cdot (C - C_{sim})$, where the scaling factor $a$ is empirically set to 2. Using these updated correlation curves, a new simulation is performed, resulting in speckle correlation dynamics that more closely match those observed in the template data.

\subsection{A simulation strategy for enhanced reference motion}
\label{sec:iterative_sim}

% Segmentation is one of the most advanced areas of research in deep learning, especially for medical imaging, and is therefore a logical place to start for developing tracking methods. 

% Let $\mathcal{D}_0 = \{ (V_1, M_1^s), ..., (V_1, M_1^s)\}$ where $V_i$ is a video and $M_i^s$ is a corresponding mesh extracted from the 2D+t segmentation map. With this initial dataset a simulated dataset $\mathcal{D}^{sim}_0 = \{ (S_1, M_1^s), ..., (S_N, M_N^s)\}$ can be generated with the pipeline metionned above. With this dataset a motion estimation model can be trained $f_0$. 

% Now given a new set of 

% Experimental analysis of validation curves, indicated high levels of overfitting for strategy 2 with version 0, to the extend that a model trained on strategy 1 would perform better on strategy 2 than a model trained on strategy 2.  For this reason, we have 

%%%%%%%%%%%

Despite the improved texture realism achieved by the previous strategies, the current simulation pipeline remains limited in terms of myocardial motion fidelity. This limitation stems from the reliance on 2D+t segmentation labels, which, through the adopted remeshing strategy, provide only a coarse approximation of true \added{local} myocardial motion. Alternative approaches have incorporated biomechanical heart models to derive more realistic motion patterns~\cite{alessandrini_tuffc_2018}. However, such methods typically rely on complex patient-specific personalization procedures, which substantially limit their scalability and practical deployment at scale. To mitigate these limitations, we introduce an innovative iterative simulation strategy inspired by the data engine of the Segment Anything Model (SAM)~\cite{kirillovSegmentAnything2023}. Following a similar philosophy, we develop a multi-stage refinement process. 

% in which simulated data and predicted temporal meshes mutually reinforce each other.

\noindent\textbf{Simulated dataset using temporal meshes from myocardial masks.} 
Let $\mathcal{V} = \{\mathbf{V}_1, \mathbf{V}_2, ..., \mathbf{V}_N\}$ denote a set of $N$ echocardiographic videos. Through either manual annotation or automated segmentation, a corresponding set of 2D+t segmentation masks is obtained, which can be converted into a collection of meshes approximating myocardial motion, denoted by $\mathcal{M}^0 = \{ \mathbf{M}_1^0, ..., \mathbf{M}_N^0\}$. Given $\mathcal{V}$, $\mathcal{M}^0$, and the pipeline in ~\cref{fig:pipeline}, an initial simulated dataset can be generated as $\mathcal{D}^0 = \{ (\mathbf{\tilde{V}}_1^0, \mathbf{M}_1^0), ..., (\mathbf{\tilde{V}}_N^0, \mathbf{M}_N^0)\}$, where each $\mathbf{\tilde{V}}_i^0$ denotes a simulated video with the same dimensions as $\mathbf{V}_i$.

\noindent\textbf{Simulated dataset using temporal meshes from a deep learning motion estimation model.} Using the initial simulated dataset $\mathcal{D}^0$, an arbitrary deep learning motion estimation model is trained. This model is applied to the real video set $\mathcal{V}$ to predict refined, less regularized, and potentially more realistic temporal meshes. For each video $\mathbf{V}_i$, only a 2D segmentation is required to initialize the tracking process, resulting in an updated set of meshes $\mathcal{M}^1 = \{ \mathbf{M}_1^1, ..., \mathbf{M}_N^1\}$. With $\mathcal{V}$ and $\mathcal{M}^1$, a new simulated dataset $\mathcal{D}^1 = \{ (\mathbf{\tilde{V}}_1^1, \mathbf{M}_1^1), ..., (\mathbf{\tilde{V}}_N^1, \mathbf{M}_N^1)\}$ is generated. This refined dataset is subsequently used to train a \replaced{new updated}{more accurate} motion estimation model.

\noindent\textbf{Multi-stage process.}
By combining an efficient segmentation model, such as %the one proposed in~
\cite{judge_miccai_2024}, with the most recently trained motion estimation model, the previous step can be iteratively repeated in a fully automated manner to progressively increase both the realism and the size of the simulated dataset. For clarity, we keep $N$ constant in the notation, although the number of videos in the dataset may vary across stages. Notably, in contrast to SAM, the predicted meshes are not manually validated at each iteration. Nevertheless, small motion estimation errors do not destabilize the proposed procedure, as the simulation step enforces exact motion in the generated images, consistent with the underlying meshes. This mechanism prevents the system from collapsing into a confirmation-bias loop, in which erroneous predictions would otherwise be progressively reinforced.

\begin{figure*}[tp]
\centering
\includegraphics[width= \textwidth]{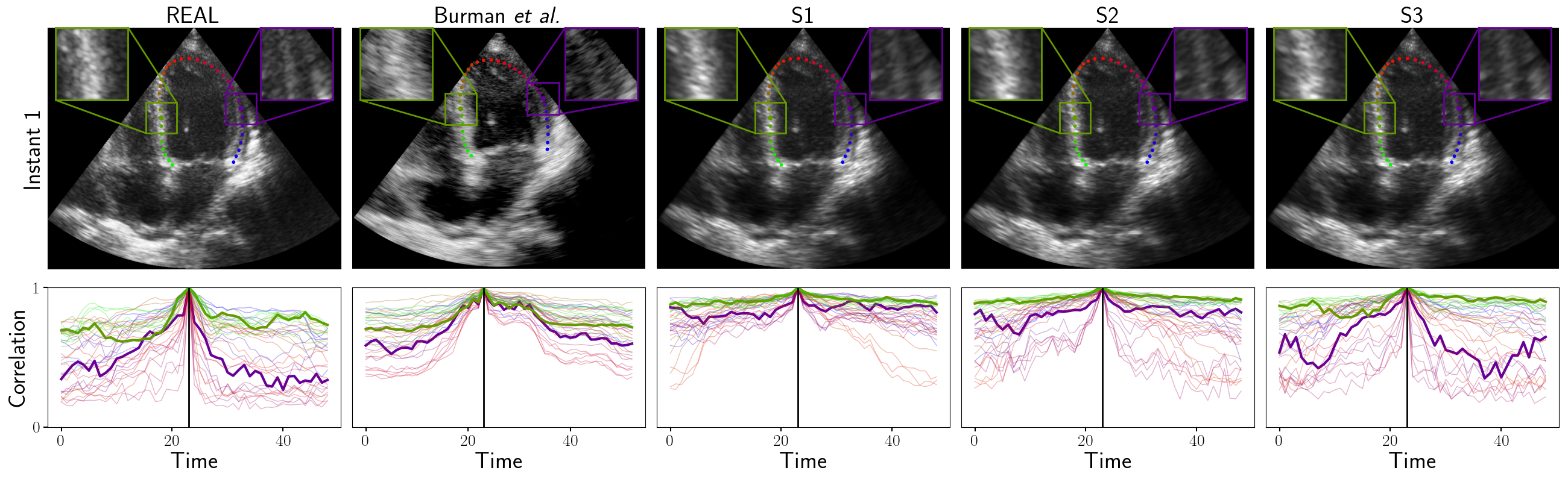}
\caption{Sample frames from a simulated sequence and its corresponding real template sequence are shown. The last row displays correlation curves computed for the 36 centerline points, with highlighted curves corresponding to points located within the zoomed image regions. Note that the simulation proposed by Burman \textit{et al.} has a different sequence length.}
% Full video sequences are provided in the supplementary material.}
\label{fig:simu_curves}
\end{figure*}

% \begin{figure}[tp]
% \centering
% \includegraphics[width=\linewidth]{figures/patient0034_col.png}
% \caption{Sample frames from a simulated sequence and its corresponding real template sequence are shown. The last row displays correlation curves computed for the 36 centerline points, with highlighted curves corresponding to points located within the zoomed image regions. Note that the simulation proposed by Burman \textit{et al.} has a different temporal length.}
% % Full video sequences are provided in the supplementary material.}
% \label{fig:simu_curves}
% \end{figure}

\subsection{Track Any Speckle Network (TAS-Net)}

The proposed synthetic dataset enables fully supervised training of deep learning models, thereby facilitating their adaptation to the temporal characteristics of echocardiographic image sequences. Among existing approaches, EchoTracker~\cite{azadEchoTrackerAdvancingMyocardial2024} represents a strong baseline; however, recent advances in joint point tracking suggest that improved performance may be achievable for this task. In particular, CoTracker-based models benefit from jointly tracking multiple points, which is especially advantageous in regions with low temporal correlation. Within this family of methods, MyoTracker~\cite{chernyshovLowComplexityPoint2025} has recently been proposed as a low-latency alternative, achieving performance close to that of CoTracker while prioritizing computational efficiency. In contrast, our work adopts the opposite design choice, prioritizing high tracking accuracy while maintaining a reasonable processing time on the order of a few seconds, consistent with clinical practice. To this end, we design a set of targeted adaptations to the CoTracker architecture, yielding an efficient model for echocardiographic video analysis. By preserving the core architecture, the proposed approach benefits from pretrained weights learned on large-scale natural video datasets, a key factor for stability and performance in our iterative training process when simulated data are scarce.

%In order to complete the iterative dataset refinement, a tracking model is required. With many models previously proposed, many choices are possible. EchoTracker is a strong candiate but recent advances in joint tracking may offer better performance for the task. A model like CoTracker has the advantage of jointly tracking points, which can be particularly helpful in low-correlation zones. Although MyoTracker makes interesting modifications for low-latency tracking, it is not the goal of our current work. Moreover, by using the CoTracker architecture without any changes, it is possible to use pre-trained weights obtained from training the model on many thousands of videos. 

%We apply different modifications to CoTracker to produce a more adapted version. Contrary to Myotracker, our goal is to obtain top performance rather than minimize the inference time. For this reason, all changes were done without affecting the core architecture of CoTracker, therefore, allowing the use of available pretrained weights, which proved essential, especially when few simulations were available. 

\subsubsection{A mesh embedding for improved spatio-temporal coherence}

% In CoTracker's default configuration, the transformer input token associated with a point $k$ at time $t$ is constructed by concatenating positional, visibility, appearance, correlation, and positional encoding features, and is defined as
In CoTracker's default configuration, the transformer input token associated with a point $k$ at time $t$ is constructed by concatenating position ($\hat{P}_t^k$), visibility ($\hat{v}_t^k$), appearance ($Q_t^k$), correlation ($C_t^k$), and positional encoding ($\eta$ and $\eta'$) features, and is defined as:
\begin{equation}
G_t^k = \Bigl(
\hat{P}_t^k - \hat{P}_1^k;
\hat{v}_t^k;
Q_t^k;
C_t^k;
\eta(\hat{P}_t^k - \hat{P}_1^k) + \eta'(\hat{P}_1^k) + \eta'(t)
\Bigr).\nonumber
\end{equation}
% where $\hat{P}_t^k$ denotes the estimated position of point $k$ at time $t$, $\hat{v}_t^k$ its visibility indicator, $Q_t^k$ an appearance vector, and $C_t^k$ correlation features extracted from the CNN feature maps. The functions $\eta$ and $\eta'$ correspond to sinusoidal positional encodings. At each iteration, the position and appearance vectors are refined using the output of the transformer module.

%At each iteration $m$, the position and appearance vectors are refined using the output of the transformer module $\Psi$, according to
%\begin{equation}
%\Delta \hat{P}, \Delta Q
%= \Psi \left(
%G \left(
%\hat{P}^{(m)}, \hat{v}^{(0)}, Q^{(m)}
%\right)
%\right).
%\end{equation}

In the default configuration, each transformer token $G_t^k$ does not encode any explicit information linking it to a specific mesh point. To address this limitation, we introduce a mesh-position embedding that is added to the appearance vector $Q_t^k$. Specifically, a time-invariant embedding $E^k$ is defined for each mesh point $k$ as $E^k = \psi(l_k, r_k)$ where $l_k$ and $r_k$ denote the normalized longitudinal and radial coordinates of the mesh point, respectively, with values in $[0,1]$. The function $\psi$ is implemented as a multilayer perceptron with three hidden layers (32, 64, and 128 neurons), with the output dimensionality matching that of the appearance vector $Q$. At each iteration, the mesh-aware appearance vector is defined as $\bar Q_t^k = Q_t^k + E^k$, and is used in the construction of the transformer input tokens. By relying on a continuous embedding conditioned on normalized mesh coordinates, the proposed formulation enables the model to be trained and applied to meshes with varying spatial resolutions and different numbers of mesh points.

\subsubsection{A continuous visibility score for explicit modeling of speckle decorrelation}

The default CoTracker model outputs not only point trajectories but also a binary visibility indicator for each point at each time frame to handle occlusions. We leverage this mechanism to explicitly incorporate speckle decorrelation into our model. Specifically, the binary visibility indicator is extended to a continuous visibility score, enabling the prediction of the degree of speckle decorrelation at each point. When trained on the proposed simulations with varying correlation levels, the network learns to regress the speckle correlation value using a mean squared error loss.

\subsubsection{A bidirectional sliding-window strategy for training and inference}

In CoTracker, points can be initialized at any time within a video and subsequently tracked across the remaining frames. However, since the myocardium is typically more \deleted{fully} visible within the ultrasound sector at the end-systolic (ES) frame \deleted{, and speckle patterns are generally clearer and more temporally consistent at this time point,} we initialize all mesh points at ES. In contrast to recent studies~\cite{taskenEstimationSegmentalLongitudinal2025}, we adopt a bidirectional tracking strategy during both training and inference. Specifically, we replace CoTracker \added{V1}’s forward-only sliding-window scheme with a bidirectional sliding-window formulation. The model is initialized at the ES frame and predicts point trajectories forward from ES to the end of the sequence, as well as backward from ES to the beginning. We retain the original window overlap strategy, whereby consecutive windows overlap by half of their extent, and predictions in the overlapping region are used to initialize or constrain the subsequent window to ensure temporal consistency.

During training, CoTracker operates on temporal crops extracted from the full sequence. In the default configuration, crops are sampled randomly and initialization occurs at the first frame of each crop. However, 
as %because 
our correlation-based prediction requires access to the ES frame to estimate relative visibility, we modify the sampling strategy to enforce that each training crop includes the ES frame. A direct consequence of always initializing at ES is a reduction in the diversity of crop–query configurations observed during training: with ES-fixed initialization, each video effectively contributes a single initialization condition, rather than multiple conditions across different crops. To preserve this diversity and improve generalization, we therefore initialize the model at an arbitrary frame during training—while still ensuring that the crop contains the ES frame. At inference, %time, 
initialization is performed exclusively at ES, in accordance with the intended clinical workflow.

\section{Experimental setup}

\begin{table}[t]
\centering
\caption{Mean trajectory error (MTE, in mm) computed for the different configurations of the proposed multi-stage strategy. Best values are in \textbf{bold}.}
\label{tab:versions}

\renewcommand{\arraystretch}{1.2}
\setlength{\tabcolsep}{6pt}
\begin{tabular}{ccccc}
\toprule
Version & Strategy & Size  & SSHF ($\downarrow$) & HUNT($\downarrow$)    \\
\midrule
$\mathcal{D}^0$ & S1                                                                     & \multirow{2}{*}{242}      & 1.21\tiny{$\pm$0.02}	& 1.42\tiny{$\pm$0.01}  \\
$\mathcal{D}^0$ & S2                                                                     &                           & 1.21\tiny{$\pm$0.04}	& 1.50\tiny{$\pm$0.04}  \\
\midrule                            
$\mathcal{D}^1$ & S1 $\rightarrow$ S1                                                    & \multirow{3}{*}{1,478}     & 1.20\tiny{$\pm$0.02}	& 1.33\tiny{$\pm$0.02} \\
$\mathcal{D}^1$ & S2 $\rightarrow$ S2                                                    &                           & \textbf{1.15\tiny{$\pm$0.01}}	& 1.37\tiny{$\pm$0.02} \\
$\mathcal{D}^1$ & S1 $\rightarrow$ S2                                                    &                           & 1.16\tiny{$\pm$0.01}	& 1.31\tiny{$\pm$0.02} \\
\midrule                            
$\mathcal{D}^2$ & S1 $\rightarrow$ S1 $\rightarrow$ S1                                   & \multirow{4}{*}{1,478}     & \textbf{1.15\tiny{$\pm$0.00}}	& 1.33\tiny{$\pm$0.01} \\
$\mathcal{D}^2$ & S2 $\rightarrow$ S2 $\rightarrow$ S2                                   &                           & \textbf{1.15\tiny{$\pm$0.00}}	& 1.34\tiny{$\pm$0.02} \\
$\mathcal{D}^2$ & S1 $\rightarrow$ S2 $\rightarrow$ S2                                   &                           & 1.16\tiny{$\pm$0.00}	& 1.32\tiny{$\pm$0.01} \\
$\mathcal{D}^2$ & S1 $\rightarrow$ S2 $\rightarrow$ S3                                   &                           & 1.16\tiny{$\pm$0.01}	&\textbf{ 1.29\tiny{$\pm$0.00}} \\
\hline
\end{tabular}
\end{table}

\subsection{Datasets used in this study}

\subsubsection{Training}

Data used in our simulation pipeline were obtained from the CAMUS and CARDINAL datasets. 
%The CAMUS dataset
CAMUS~\cite{leclercDeepLearningSegmentation2019} comprises apical two-chamber (A2C) and four-chamber (A4C) views from 500 patients. Manual annotations of the LV endocardium and epicardium, as well as the left atrium, were performed by a senior cardiologist for both end-diastolic (ED) and end-systolic (ES) frames. In addition, a subset of 98 4CH videos was manually annotated over the entire cardiac cycle~\cite{painchaudEchocardiographySegmentationEnforced2022}. %The CARDINAL dataset
CARDINAL~\cite{lingExtractionVolumetricIndices2023} includes 239 hypertensive patients, for whom pseudo-labels were generated using a 2D+t nnU-Net for both 2CH and 4CH acquisitions and validated by an expert cardiologist. 
% Across both datasets, video sequences have an average length of 48$\pm$11, with spatial dimensions varying from 323 to 915 pixels in width and from 292 to 778 pixels in height. The two datasets have the same spatial resolution of $0.3$ mm$^2$. 

In addition, we independently used the simulated dataset proposed by Burman \textit{et al.}~\cite{burmanLargescaleSimulationRealistic2024} described in Section \ref{sec:related}. 
% This dataset comprises 1,296 synthetic echocardiographic sequences generated from 50 base patients of the CAMUS dataset. 
% The videos have an average length of 50$\pm$15 frames, with spatial resolutions ranging from 383 to 874 pixels in width and from 316 to 631 pixels in height. 
Reference motion was extracted from the available scatterer displacements and used to derive reference temporal meshes, consistent with our simulation pipeline. 

%Data for our simulation pipeline was obtained from the CAMUS and CARDINAL datasets. The CAMUS dataset~\cite{leclercDeepLearningSegmentation2019} consists of apical 2-chamber (2CH) and 4-chamber (4CH) views of 500 patients. Manual annotations of the left ventricular endocardium and epicardium, along with the atrium, were performed by a senior cardiologist for both ED and ES images. A subset of 98 4CH videos was also labeled for the entire cardiac cycle~\cite{painchaudEchocardiographySegmentationEnforced2022}. The CARDINAL dataset~\cite{lingExtractionVolumetricIndices2023} consists of 239 hypertensive patients for which pseudo labels were generated by a 2D+t nnU-Net for both the 2CH and 4CH acquisitions. 

% This dataset has video that have an averange length of 46 frames (15 to 102) and size ranging between 323 to 915 pixels in width and 292 to 778 pixels in height. 

% In addition, we separately used the simulated dataset created by Burman \textit{et al.}~\cite{burmanLargescaleSimulationRealistic2024}. This dataset contains 1296 sequences simulated from 50 base patients in the CAMUS dataset. Videos have 50 frames on average (27 to 100) and range in size between 383 and 874 pixels in width and 316 to 631 pixels in height. 

\subsubsection{Evaluation}

Three complementary datasets were used for evaluation. Two of them, SSHF and HUNT~\cite{HUNT_SSHF}, were specifically designed for test–retest analysis and include 40 and 30 patients, respectively. For each patient, two independent recordings were acquired in the three standard apical views (A2C, A3C, and A4C), and each recording was independently analyzed by a different expert. 
% This dataset allows the evaluation of the agreement between experts while considering the probe position, the cardiac cycle selection, and the tracking initilization. 
The third dataset, HUNT-A, comprises 88 patients~\cite{HUNTA}. For each patient, a single recording was available for each of the three apical views, and each recording was independently analyzed by two experts. 
SSHF and HUNT enable evaluation of inter-expert agreement while accounting for probe position, cardiac-cycle selection, and tracking initialization, whereas HUNT-A evaluates inter-expert agreement without probe-position variability.

% This dataset allows the evaluation of the agreement between experts without considering the position of the probe. 

All videos were semi-automatically annotated \added{by clinical experts} using the EchoPac commercial software, which performs speckle tracking to extract myocardial points tracked over the entire cardiac cycle. The resulting annotations consist of a set of $K$ two-dimensional points of size $T \times K \times 2$, representing the temporal tracking of the myocardial centerline. These points are subdivided into six segments for regional strain analysis. EchoPac automatically proposes 
to accept or reject %the acceptance or rejection of
each segment based on internal
%, 
proprietary criteria, which can be manually overridden by the user. The proportions of accepted segments were 92\%, 82\%, and 92\% for SSHF, HUNT, and HUNT-A. 
% On average, videos contain 89$\pm$19 frames. 
The myocardial centerline is represented by an average of 73 points, with a mean inter-point spacing of 2.5 mm. Reference global longitudinal strain (GLS) values were computed directly from the extracted tracking points, rather than using the GLS estimates provided by EchoPac, as the latter rely on proprietary and undisclosed computation procedures. Across all datasets, the mean peak GLS was -17.8$\pm$3.2\%.

\subsection{Track Any Speckle dataset}

Our dataset, TAS-1K, comprises 1,478 synthetic echocardiographic videos, totaling 69,535 frames, along with corresponding reference myocardial motion encoded as temporal meshes. 
% We compare TAS-1K with the dataset introduced by Burman \textit{et al.}, which consists of 1,296 synthetic videos, and analyze the impact of both datasets on the accuracy of deep learning–based motion estimation methods trained on them. 
% TAS-1K is released to support future research and development in myocardial motion estimation from echocardiographic videos.
This simulated dataset was generated using the methodology described in Section~\ref{sec:iterative_sim}. The first stage relied on 242 real echocardiographic sequences drawn from the CAMUS (98 \replaced{videos}{cases}) and CARDINAL (144 \replaced{videos}{cases}) datasets.
% acquired at two hospitals in France (University Hospital of St Etienne and Hospices Civils of Lyon). 
These cases were reviewed by expert clinicians to validate the accuracy of the 2D+t segmentation masks, which were obtained either through manual annotation (CAMUS) or via an automatic segmentation model (CARDINAL)~\cite{lingExtractionVolumetricIndices2023}. This process enables the generation of the initial synthetic dataset $\mathcal{D}^0$.
% $\mathcal{D}^0 = \{ (\mathbf{\tilde{V}}_1^0, \mathbf{M}_1^0), ..., (\mathbf{\tilde{V}}_{242}^0, \mathbf{M}_{242}^0)\}$.

The second stage relied on 1,478 real echocardiographic videos drawn from the full CAMUS (1,000 \replaced{videos}{cases}) and CARDINAL (478 \replaced{videos}{cases}) datasets. For all cases, the mesh was initialized at the ES frame, using either the manual annotations %provided with the 
(CAMUS) 
or segmentations obtained from an automatic method validated by an expert clinician %for the 
(CARDINAL)\cite{lingExtractionVolumetricIndices2023}. Inspired by the SAM framework, TAS-Net was trained on $\mathcal{D}^0$ to estimate myocardial motion for these 1,478 \replaced{videos}{cases}. This process enabled the generation of the 
$\mathcal{D}^1$ 
dataset
and was repeated once to generate 
the 
$\mathcal{D}^2$ 
dataset, %datasets, 
each comprising 1,478 synthetic videos. 
% The \mbox{TAS-Net} model trained on $\mathcal{D}^2$ was used as the optimal model in the experimental section. 

Our refinement framework is flexible and supports multiple configurations, particularly with respect to the choice of simulation strategies used to model speckle decorrelation, as described in Section~\ref{sec:speckle_decorrelation_strategies}. We therefore evaluated several configurations, summarized in Table~\ref{tab:versions}. For example, \mbox{S1 $\rightarrow$ S1 $\rightarrow$ S1} denotes a three-stage refinement process in which the same speckle decorrelation strategy S1 is applied at all stages, whereas \mbox{S1 $\rightarrow$ S2 $\rightarrow$ S3} corresponds to a progressive refinement scheme using increasingly realistic speckle decorrelation strategies across stages.

\subsection{Model configuration}

Three point-tracking models were assessed in this study.

\noindent\textbf{EchoTracker.} We evaluate both the EchoTracker model released with the original publication and a variant retrained on simulated data. The released model was trained on 6,490 real A2C, A3C, and A4C echocardiographic videos.
% Following the original training protocol, we retrained EchoTracker on simulated data using a batch size of 1, a learning rate of $5 \times 10^{-4}$, a one-cycle learning rate scheduler, and the AdamW optimizer. Using a batch size of 1 removes constraints on sequence length, as full video sequences fit within GPU memory. 
Following the original training protocol, we retrained EchoTracker on simulated data.
We additionally evaluated an initialization at the end-systolic (ES) frame; however, this configuration resulted in degraded performance compared with initialization at the first frame, as in the original model.

%We evaluate both the EchoTracker model released with the original paper and an EchoTracker model trained on simulated data. The released model was trained on real data comprising A4C, A2C and A3C subsets with a total of 6,490 videos. Following the original training procedure, we trained Echotracker on simulated data using a batch size of 1, a learning rate of $5 \cdot 10^{-4}$, a one-cycle scheduler~\cite{onecycle}, and the AdamW optimizer~\cite{adam}. Using a batch size of 1 removed restrictions on sequence length, as full sequences fit in GPU memory. We also tested initializing tracking at the ES frame, but this yielded worse results than initializing at the first frame, as in the original model.

\noindent\textbf{MyoTracker.} MyoTracker was implemented as described in~\cite{chernyshovLowComplexityPoint2025}. No noticeable performance difference was observed between full-resolution inputs and resized $(256 \times 256)$ images; therefore, the latter configuration was used. 
% Tracking was initialized at the ES frame, which improved accuracy. Training employed the AdamW optimizer with an exponential decay scheduler ($\gamma = 0.99995$) and an initial learning rate of $10^{-3}$. A fixed sequence length of 64 frames was used, with shorter sequences padded using their reversed versions. Reintroducing the iterative refinement step further improved performance.
Tracking was initialized at the ES frame, which improved accuracy, as did reintroducing the iterative refinement steps.

% We implemented MyoTracker as described in~\cite{chernyshovLowComplexityPoint2025}. We tested MyoTracker at both full resolution and a resized resolution of $256 \times 256$ pixels and observed no consistent advantage for either setting. We therefore used the $256 \times 256$ pixel resolution, as described in the paper. Because the model does not use a sliding window, we initialized tracking at the ES frame, which improved tracking accuracy. We used the AdamW optimizer with an exponential decay scheduler ($\gamma = 0.99995$) and an initial learning rate of $10^{-3}$. A sequence length of 64, padding the video with its reverse when the sequence length is too short. We also found that we obtained better results when re-adding the iterative refinement, originally removed from the CoTracker framework for improved processing times. 

\noindent\textbf{TAS-Net.} TAS-Net was built upon the CoTracker1 architecture (CoTracker1 outperformed CoTracker3 in initial tests). TAS-Net was trained with the AdamW optimizer and a one-cycle learning rate scheduler using an initial learning rate of $5 \times 10^{-4}$. Training was performed with a batch size of 1 and sequence lengths up to 24 frames, without padding, using full-resolution images. A sliding-window length of 8 frames was adopted, consistent with CoTracker. All experiments were initialized from publicly available pretrained checkpoints.

% We trained TAS-Net using the AdamW optimizer with a one-cycle scheduler and a learning rate of $5 \cdot 10^{-4}$. We used a batch size of 1 and sequences of length 24 or less (no padding). We used the full image resolution for both training and testing. The sliding window length was 8, as in the CoTracker paper. In all training runs, we initialized model weights from publicly available pre-trained checkpoints.

All models were trained for 200,000 optimization steps, and the best-performing checkpoint was selected based on the validation loss. Each model was trained on both our simulated dataset and the dataset introduced by Burman \textit{et al.} In both cases, temporal data augmentation was applied by uniformly skipping one to three frames to emulate faster motion and variable frame rates. At inference time, EchoTracker and \mbox{MyoTracker} process test-set frames in $3.8\pm0.9$ ms and $3.1\pm0.6$ ms, respectively. TAS-Net processes frames in $26.7\pm3.7$ ms, which is longer but remains compatible with clinical practice. 
% At inference time, EchoTracker and MyoTracker process test-set sequences in $0.27\pm0.96$ s and $0.13\pm0.05$ s, respectively, (0.002 and 0.004 s per frame). TAS-Net processes sequences in $2.32\pm0.52$ s (0.03 s per frame), which is longer but remains compatible with clinical practice. 
% Runtime variability arises from differences in image resolution and in the number of query points. 
All runtimes are reported for an NVIDIA A40 GPU with 48 GB of memory. During inference, a mesh with $r=5$ radial points is generated around the centerline for MyoTracker and \mbox{TAS-Net}, as these methods benefit from additional support points through joint tracking. In contrast, EchoTracker performs independent point tracking and therefore does not benefit from supplementary mesh points.

% All models were trained for 200,000 steps, and the best checkpoint was selected according to the validation loss. We trained all three models on both our simulated dataset ($\mathcal{D}^2_{S1 \shortto S2 \shortto S3}$) and the Burman \textit{et al.} dataset. In both cases, we used temporal data augmentation by skipping every 1 to 3 frames to mimic fast motion and frame rates. 

% EchoTracker and MyoTracker process test-set sequences in $0.27\pm0.96$ s and $0.13\pm0.05$ s, respectively (0.002 and 0.004 s per frame). CoTracker processes sequences in $2.32\pm0.52$ s (0.03 s per frame), which is longer but still acceptable in clinical practice. Runtimes vary due to differences in image size and the number of query points. All times are reported for an NVIDIA A40 GPU with 48 GB of memory.

% During inference, we generate a mesh with $r=5$ around the centerline points for MyoTracker and CoTracker as they benefit from support points due to joint tracking. EchoTracker tracks points individually and does not benefit from supplementary points. 

%\subsection{Synthetic dataset generation}

\begin{table}[tp]
\centering
\small
\caption{Ablation study of the main contributions introduced in the design of \mbox{TAS-Net}. 
% Bidirectional sliding window (Bi-window), continuous visibility loss (Vis. loss), and mesh embedding are evaluated in terms of mean trajectory error (MTE), GLS mean absolute error (MAE), and mesh folding (Folds). 
Row 1 corresponds to the baseline model described in\cite{taskenEstimationSegmentalLongitudinal2025}. Best values are in \textbf{bold}.
}
% \begin{tabular*}{\textwidth}{ccc @{\extracolsep{\fill}} ccc ccc}
% \toprule
% \multicolumn{3}{c}{Method} & \multicolumn{3}{c}{SSHF - 240 videos ($\downarrow$)} & \multicolumn{3}{c}{HUNT - 180 videos ($\downarrow$)}  \\
% \cmidrule(lr){1-3} \cmidrule(lr){4-6}  \cmidrule(lr){7-9} 

%  \mcc{Bi-window} & \mcc{Vis. loss} & \mcc{Mesh embedding} &  \mcc{MTE (mm)}  & \mcc{GLS MAE (\%)} & \mcc{Folds (\%)} &  \mcc{MTE (mm)}  & \mcc{GLS MAE (\%)} & \mcc{Folds (\%)}   \\
% \midrule

% \ding{55}     & \ding{55}     & \ding{55}     &1.18\tiny{$\pm$0.00}	& 1.46\tiny{$\pm$0.06}		& 27\tiny{$\pm$2}		& 1.34\tiny{$\pm$0.01}		& 2.20\tiny{$\pm$0.12}		& 42\tiny{$\pm$8} \\
% \checkmark    & \ding{55}     & \ding{55}     &1.17\tiny{$\pm$0.00}	& 1.52\tiny{$\pm$0.01}		& 22\tiny{$\pm$1}		& 1.32\tiny{$\pm$0.00}		& 2.33\tiny{$\pm$0.12}		& 32\tiny{$\pm$4} \\
% \checkmark    & \checkmark    & \ding{55}     &1.16\tiny{$\pm$0.01}	& 1.42\tiny{$\pm$0.03}		& 25\tiny{$\pm$1}		& 1.29\tiny{$\pm$0.00}		& 2.03\tiny{$\pm$0.01}		& 33\tiny{$\pm$2} \\
% \checkmark    & \checkmark    & \checkmark    &\textbf{1.15\tiny{$\pm$0.01}}	& \textbf{1.32\tiny{$\pm$0.01}}		& \textbf{18\tiny{$\pm$1}}		& \textbf{1.28\tiny{$\pm$0.01}}		& \textbf{1.84\tiny{$\pm$0.05}}		& \textbf{18\tiny{$\pm$1}} \\

% \bottomrule      
% \end{tabular*}

\begin{tabular}{ccc  ccc}
\toprule
\multicolumn{3}{c}{Method} & \multicolumn{3}{c}{SSHF - 240 videos ($\downarrow$)}  \\
\cmidrule(lr){1-3} \cmidrule(lr){4-6} 

 \mcc{\makecell{Bi- \\ window}} & \mcc{\makecell{Vis. \\ loss}} & \mcc{\makecell{Mesh\\embedding}} &  \mcc{\makecell{MTE \\ (mm)}}  & \mcc{\makecell{GLS MAE \\ (\%)}} & \mcc{\makecell{Folds \\ (\%)}}    \\
\midrule

\ding{55}     & \ding{55}     & \ding{55}     &1.18\tiny{$\pm$0.00}	& 1.46\tiny{$\pm$0.06}		& 27\tiny{$\pm$2}		 \\
\checkmark    & \ding{55}     & \ding{55}     &1.17\tiny{$\pm$0.00}	& 1.52\tiny{$\pm$0.01}		& 22\tiny{$\pm$1}		 \\
\checkmark    & \checkmark    & \ding{55}     &1.16\tiny{$\pm$0.01}	& 1.42\tiny{$\pm$0.03}		& 25\tiny{$\pm$1}		 \\
\checkmark    & \checkmark    & \checkmark    &\textbf{1.15\tiny{$\pm$0.01}}	& \textbf{1.32\tiny{$\pm$0.01}}		& \textbf{18\tiny{$\pm$1}}		 \\

\bottomrule      
\end{tabular}
\label{tab:ablation}
\end{table}

\begin{figure}[tp]
    \centering
    \includegraphics[width=0.9\linewidth]{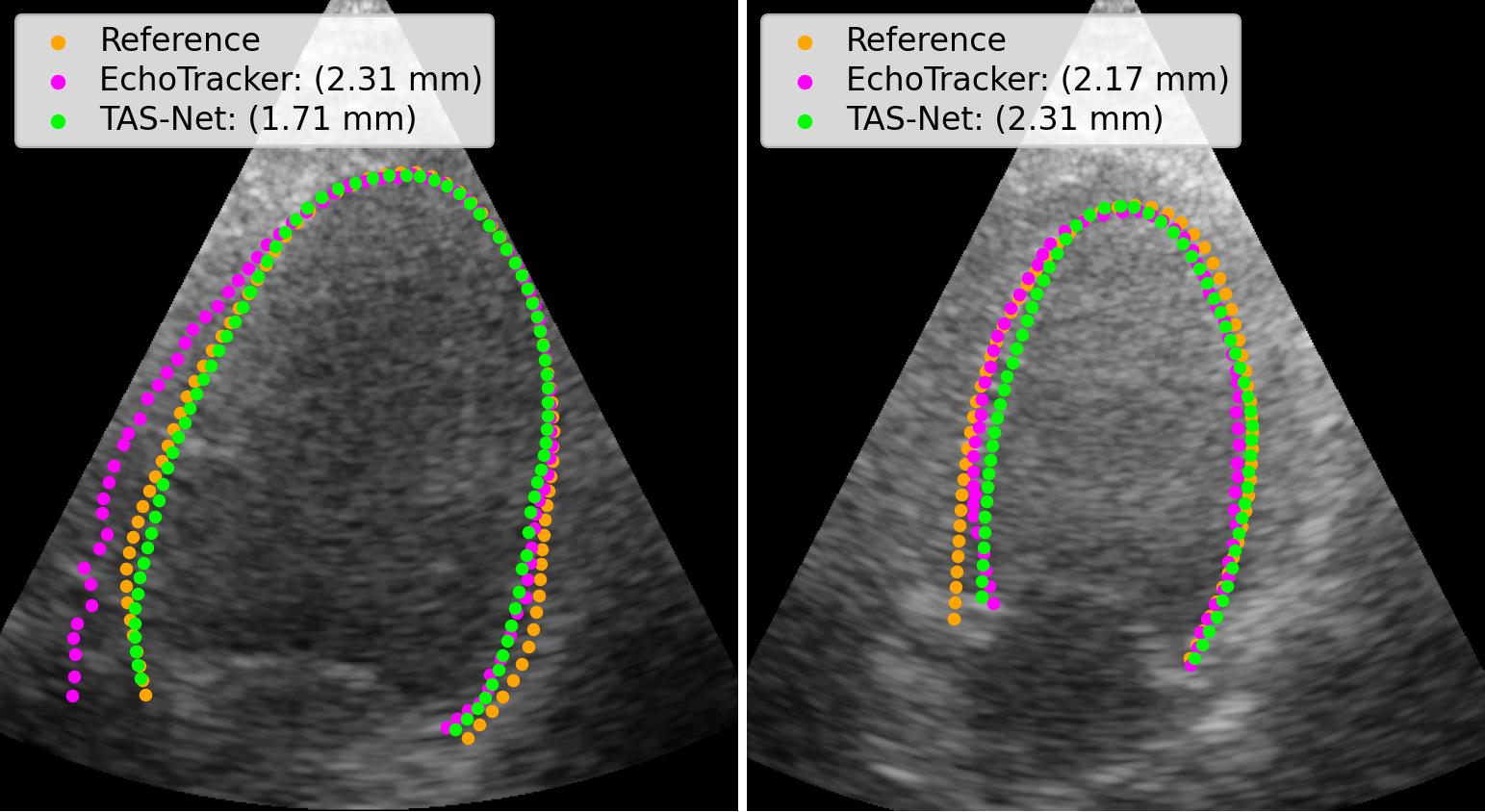}
    % \caption{Tracking curves obtained with \mbox{TAS-Net} and EchoTracker trained on real images for their respective worst-case sequences from the HUNT dataset. Displayed values correspond to the mean trajectory error (mm).}
    \caption{Tracking results on respective worst-case sequences from the HUNT dataset for EchoTracker trained on real images [left] and \mbox{TAS-Net} [right], with their mean trajectory error.}
    % Full videos are available in the supplementary material.}
    \label{fig:worst_sample}
\end{figure}

\subsection{Evaluation}

\begin{table*}[tp]
\centering
\caption{Mean trajectory error (MTE, mm) across datasets and views for the evaluated methods. Best scores are indicated in \textbf{bold} for methods trained on synthetic data.} 
\label{tab:sota_geometric}
\small
% \begin{tabular*}{\textwidth}{lll@{\extracolsep{\fill}} cccccc}
\begin{tabular*}{\textwidth}{lc @{\extracolsep{\fill}} ccc ccc ccc}
% \begin{tabular*}{\textwidth}{@{\hspace{0pt}}l cccccc cccccc@{\hspace{0pt}}}
\toprule
% \multicolumn{4}{c}{Method} & \multicolumn{3}{c}{SSHF} & \multicolumn{3}{c}{HUNT}  \\
\multirow{2}{*}{{Method}} & \multirow{2}{*}{{Dataset}} &  \multicolumn{3}{c}{{SSHF - 240 videos ($\downarrow$)}}  & \multicolumn{3}{c}{{HUNT - 180 videos ($\downarrow$)}} & \multicolumn{3}{c}{{HUNT-A - 588 videos ($\downarrow$)}} \\
 \cmidrule(lr){3-5} \cmidrule(lr){6-8} \cmidrule(lr){9-11}

& & A2C & A3C & A4C & A2C & A3C & A4C & A2C & A3C & A4C \\
\midrule

% \rowcolor{gray!15}
EchoTracker      & Real                                             & 1.21&					1.09&					1.10&					1.45&					1.32&					1.27&					1.26&					1.15&					1.04	\\
\midrule

EchoTracker      & Burman     & 2.30\tiny{$\pm$0.05}&	2.18\tiny{$\pm$0.04}&	2.26\tiny{$\pm$0.05}&	3.28\tiny{$\pm$0.05}&	2.64\tiny{$\pm$0.01}&	2.99\tiny{$\pm$0.05}&	3.41\tiny{$\pm$0.04}&	2.63\tiny{$\pm$0.02}&	2.86\tiny{$\pm$0.03} \\
MyoTracker       & Burman     & 2.42\tiny{$\pm$0.06}&	2.59\tiny{$\pm$0.09}&	2.56\tiny{$\pm$0.06}&	3.35\tiny{$\pm$0.14}&	2.70\tiny{$\pm$0.07}&	2.99\tiny{$\pm$0.08}&	3.26\tiny{$\pm$0.06}&	2.77\tiny{$\pm$0.08}&	3.03\tiny{$\pm$0.09} \\
TAS-Net          & Burman     & 1.71\tiny{$\pm$0.08}&	1.54\tiny{$\pm$0.07}&	1.49\tiny{$\pm$0.07}&	2.54\tiny{$\pm$0.05}&	1.96\tiny{$\pm$0.10}&	2.08\tiny{$\pm$0.10}&	2.33\tiny{$\pm$0.09}&	1.90\tiny{$\pm$0.15}&	1.95\tiny{$\pm$0.11} \\
EchoTracker      & TAS-1K     & 1.57\tiny{$\pm$0.04}&	1.56\tiny{$\pm$0.07}&	1.49\tiny{$\pm$0.02}&	2.07\tiny{$\pm$0.05}&	2.02\tiny{$\pm$0.07}&	1.79\tiny{$\pm$0.05}&	1.93\tiny{$\pm$0.04}&	1.98\tiny{$\pm$0.04}&	1.68\tiny{$\pm$0.03} \\
MyoTracker       & TAS-1K     & 1.49\tiny{$\pm$0.01}&	1.49\tiny{$\pm$0.03}&	1.43\tiny{$\pm$0.06}&	1.80\tiny{$\pm$0.05}&	1.79\tiny{$\pm$0.08}&	1.49\tiny{$\pm$0.05}&	1.68\tiny{$\pm$0.05}&	1.71\tiny{$\pm$0.05}&	1.40\tiny{$\pm$0.05} \\
TAS-Net          & TAS-1K     & \textbf{1.22\tiny{$\pm$0.00}}&	\textbf{1.12\tiny{$\pm$0.00}}&	\textbf{1.12\tiny{$\pm$0.01}}&	\textbf{1.42\tiny{$\pm$0.01}}&	\textbf{1.19\tiny{$\pm$0.00}}&	\textbf{1.21\tiny{$\pm$0.01}}&	\textbf{1.33\tiny{$\pm$0.00}}&	\textbf{1.17\tiny{$\pm$0.01}}&	\textbf{1.07\tiny{$\pm$0.00}} \\

\bottomrule      
\end{tabular*}

\end{table*}

All methods were evaluated against a reference derived from the EchoPac software output, which was generated and validated by expert clinicians. Performance was assessed in terms of both geometric accuracy and application-oriented criteria by analyzing the resulting strain measurements. Only myocardial segments that were approved by expert clinicians were included in the evaluation. For all methods, we trained 3 models with different random seeds. We report mean and standard deviation values in all tables except for the strain, where one model is used, and statistical significance is reported.

\subsubsection{Motion estimation metrics}

%Motion estimation accuracy was quantified using the mean trajectory error (MTE), computed 
We computed the mean trajectory error (MTE) 
between the tracked points and the corresponding reference points. 
% We 
% also %additionally 
% assessed potential mesh folding by counting the number of instances in which the Jacobian determinant of the transformation from the initial frame ($t=0$) to time $t$ was negative for each \added{triangle} mesh cell. 
We also assessed potential mesh folding by counting the number of triangle mesh cells in which the Jacobian determinant of the transformation from the initial frame ($t=0$) to time $t$ was negative. We report the percentage of videos with at least one fold. 
% It is worth noting that the proportion of mesh folding is highly dependent on the mesh construction, including the spatial spacing between points and the orientation of the radial segments. Although this metric is relevant for evaluating the spatial consistency of the estimated motion, its applicability is limited to TAS-Net and MyoTracker, as EchoTracker only processes a myocardial centerline. For this reason, we restrict the use of this metric to the ablation study of our model.
It is worth noting that the proportion of mesh folding is in part dependent on the mesh construction, including point spacing and radial segment orientation. While useful for assessing the spatial consistency of the estimated motion, it applies only to TAS-Net and MyoTracker, as EchoTracker only processes a myocardial centerline, and is therefore only used in our ablation study.
\subsubsection{Clinical metrics}

We evaluated the prediction of peak global longitudinal strain (GLS) for all methods by computing the mean absolute error (MAE) and performing a Bland–Altman analysis with respect to the reference values. In addition, a qualitative analysis of regional strain was conducted for the best-performing methods.

% All methods were evaluated with respect to the reference, which is the output of EchoPac software, validated by clinicians. We evaluated the methods in terms of geometric accuracy and using an application-oriented evaluation by examining strain values. 

% Geometric accuracy was measured with the mean trajectory error (MTE) for the tracked points with respect to the reference points. We only considered the segments that had been approved by the clinicians in the average. 
 
% Finally, we monitored the potential folding in the tracked mesh by counting the number of times the Jacobian of the transformation of each mesh cell from time 0 to time t was negative (Lthe Jacobian was computed at each triangular cell using the RVMEP library [https://github.com/gbernardino/rvmep]). It is worth noting that the proportion of mesh folding can be highly dependent on the mesh construction (space between points, radial segment orientation). Despite its high relevance to evaluate the spatial consistency of the tracking, its relevance for our evaluation data is moderate, in particular given the reference data only consists of the myocardial centerline.

\section{Results} 

\subsection{Temporal realism of simulated ultrasound texture}

\label{sec:res_curves}

% \begin{table}[t]
%     \centering
%     \caption{MAE and Pearson correlation between speckle-correlation curves from synthetic and matched real template sequences are reported for the three simulation strategies on 98 CAMUS patients (mean ± standard deviation).}
%     \label{tab:correlation_curves}
%     \begin{tabular}{ccc}
%     \toprule
%     Strategy &  MAE ($\downarrow$) & $\rho$ ($\uparrow$) \\
%     \midrule
%     S1 &  $0.149\pm0.051$ & $0.839\pm0.066$ \\
%     S2 &  $0.119\pm0.029$ & $0.881\pm0.047$ \\
%     S3 &  $\mathbf{0.077\pm0.013}$ & $\mathbf{0.941\pm0.022}$ \\
%     \bottomrule
%     \end{tabular}

% \end{table}

We first assess the realism of each simulation strategy in terms of speckle decorrelation. This analysis was conducted on the 98 patients from CAMUS  with available ground-truth segmentations. Each sequence was simulated using the three strategies described in Section~\ref{sec:speckle_decorrelation_strategies}, without any refinement stage. Pointwise correlations between the reference ES frame and each subsequent frame were computed at every mesh location. These correlations were then compared with the corresponding values computed from the real sequences. Quantitative comparisons were performed using the mean absolute error (MAE). The S1, S2 and S3 strategies had MAEs of $0.149\pm0.051$, $0.119\pm0.029$ and $0.077\pm0.013$ respectively.

% and the Pearson correlation coefficient ($\rho$), with results reported in \Cref{tab:correlation_curves}. 

The results indicate that strategies S2 and S3 substantially improve speckle decorrelation modeling compared with S1, with S3 yielding the closest match to real data. Qualitative results are illustrated in \Cref{fig:simu_curves} for the three proposed strategies as well as for the dataset of Burman \textit{et al.} Since the latter is derived from a different subset of CAMUS patients, a strict quantitative comparison is not possible. The figure displays the correlation curves for the 36 myocardial centerline points and illustrates the increased diversity and closer agreement of speckle decorrelation patterns obtained with strategies S2 and S3 compared with the real data.

% We first start by analyzing the realism of each simulation strategy in terms of speckle decorrelation. We conduct this analysis on the 98 patients with ground truth segmentation from the CAMUS dataset. 

% We simulated each sequence with the three strategies detailed in section~\ref{sec:speckle_decorrelation_strategies} and computed the correlations at each point of the mesh and each instant. We compare these point-wise correlations with the corresponding values computed on the real sequence. We do this comparison with the mean absolute error (MAE) and Pearson correlation coefficient ($\rho$). The values are reported in \Cref{tab:correlation_curves}. 

%The values indicate that the S2 and S3 strategies offer substantial improvement over the S1 strategy. Qualitative results are shown in \Cref{fig:simu_curves} for our three strategies and Burman \textit{et al.}, which is also derived from the CAMUS dataset but uses a different patient subset, which prevents a strict quantitative comparison. The figure shows the correlation curves for the 36 centerline points.

% We first conducted a study of the 98 patients to assess the realism of the correlation curves of the generated simulations. The S1, S2 and S3 strategies had MAEs of $0.149\pm0.051$, $0.119\pm0.029$ and $0.077\pm0.013$ respectively. 

\subsection{Effectiveness of the multi-stage simulation strategy}
\label{sec:resMulti-stage Simulation Strategy_version_strategy}

\Cref{tab:versions} reports the performance of \mbox{TAS-Net} on SSHF and HUNT for different configurations of the refinement strategy, considering both the number of refinement stages and the simulation strategies used to model speckle decorrelation. First, we observe that performance generally improves with the number of refinement stages, particularly on HUNT. However, using simulation strategy S2 without refinement leads to inferior performance, which can be attributed to overfitting when training on a small dataset (242 samples). % with increased speckle realism. 
Similarly, S2 $\rightarrow$ S2 underperforms compared with S1 $\rightarrow$ S1. This can be explained by the fact that suboptimal motion estimates obtained at the first stage propagate to subsequent stages, limiting overall performance. Based on these observations, we propose a multi-stage strategy in which the realism of speckle decorrelation is progressively increased across stages, resulting in increasingly challenging training datasets. This motivates the \mbox{S1 $\rightarrow$ S2 $\rightarrow$ S3} configuration, which achieves the best overall performance across SSHF and HUNT. Based on these observations, we retain this best-performing configuration for the remaining experiments. The corresponding simulated dataset is hereafter referred to as \mbox{TAS-1K}.

% Performance of TAS-Net trained with different strategies and at different version of the iterative refinement process are presented in \Cref{tab:versions}. It can be observed that performance generatlly increase with the versions. 

% However, as observed in validation data, training with the more complex S2 strategy offers a worse performance that the S1 strategy during early stages of the refinement process due to the dataset being too small and resulting overfitting. Moreover, this performance drop cannot be compensated for completely by increasing for the data as the model used in the refinement process is sub-optimal. For example, S1 $\rightarrow$ S1 $\rightarrow$ S1 performs better than S2 $\rightarrow$ S2 $\rightarrow$ S2.  

% $\mathcal{D}^2_{S1 \rightarrow S1 \rightarrow S1}$ and $\mathcal{D}^2_{S2 \shortto S2 \shortto S2}$

% The results show that best results are obtained when slowly increasing the level of speckle decorrelation, i.e. difficulty, as the dataset gets larger and the simulated motion is more realistic. The final dataset, $\mathcal{D}^2$ and strategy S1 $\rightarrow$ S2 $\rightarrow$ S3, is used for the remaining experiments and is hereafter referred to as \mbox{TAS-1K}.%labeled $\mathcal{D}^2_{S1 \shortto S2 \shortto S3}$.

\subsection{Ablation study of the proposed TAS-Net model}

\begin{table*}[tp]
\centering

\caption{Peak GLS estimation performance, mean absolute error (MAE) and Bland–Altman statistics ($\mu$ and $\sigma$), with respect to the reference and in a test–retest/agreement setting. 
% Best values (lower) are indicated in \textbf{bold} for methods trained on synthetic data. * indicates statistically significant difference from all other methods trained on synthetic data for the MAE metric (paired one-tailed t-test, $p \geq 0.05$).
For methods trained on synthetic data, best values are indicated in \textbf{bold} and * indicates statistically significant difference from all other methods for the MAE metric (paired one-tailed t-test, $p \leq 0.05$).
} 
\label{tab:gls}

\small
% \begin{tabular*}{\textwidth}{lll@{\extracolsep{\fill}} cccccc}
\begin{tabular*}{\textwidth}{@{\extracolsep{\fill}}lc ccc ccc ccc ccc}

\toprule
% \multicolumn{4}{c}{Method} & \multicolumn{3}{c}{SSHF} & \multicolumn{3}{c}{HUNT}  \\
\multirow{3}{*}{{Method}} & \multirow{3}{*}{{Dataset}} &  \multicolumn{6}{c}{{ SSHF+HUNT  ($\downarrow$) } }  & \multicolumn{6}{c}{{HUNT-A  ($\downarrow$)}} \\
\cmidrule(lr){3-8} \cmidrule(lr){9-14}

& & \multicolumn{3}{c}{Reference - 420 videos} & \multicolumn{3}{c}{Test-retest - 210 videos} & \multicolumn{3}{c}{Reference - 588 videos} & \multicolumn{3}{c}{Agreement - 264 videos} \\
% \cline{3-14}
\cmidrule(lr){3-5} \cmidrule(lr){6-8} \cmidrule(lr){9-11} \cmidrule(lr){12-14} 
% & & MAE ($\downarrow$) & $\mu$ ($\downarrow$) & $\sigma$ ($\downarrow$) & MAE ($\downarrow$) & $\mu$ ($\downarrow$) & $\sigma$ ($\downarrow$) & MAE ($\downarrow$) & $\mu$ ($\downarrow$) & $\sigma$ ($\downarrow$) & MAE ($\downarrow$) & $\mu$ ($\downarrow$)  &$\sigma$ ($\downarrow$) \\

& & MAE  & $\mu$  & $\sigma$  & MAE  & $\mu$  & $\sigma$  & MAE  & $\mu$  & $\sigma$  & MAE  & $\mu$  & $\sigma$  \\

\midrule

%%%%%%% RESULTS WITH MYOTRACKER sequence length 36 
% \multicolumn{2}{c}{\textcolor{red}{Reference}} & -  &	-&	-&	2.02&	0.10&	2.77&	-&	-&	-&	1.78&	-0.78&	2.15  \\
% EchoTracker      & Real                        & 1.59&	0.59&	1.94&	1.98&	0.03&	2.57&	1.50&	0.95&	1.85&	1.14&	-0.34&	1.47  \\            
% \midrule
% EchoTracker      & Burman                      & 5.78&	5.72&	3.02&	2.96&	-0.73&	3.82&	6.21&	6.18&	3.28&	2.27&	-0.87&	2.99  \\            
% MyoTracker       & Burman                      & 2.69&	1.37&	3.20&	3.03&	-0.62&	3.93&	2.52&	1.09&	3.12&	2.06&	-0.39&	2.83  \\            
% TAS-Net          & Burman                      & 2.60&	1.36&	3.00&	2.54&	-0.30&	3.34&	3.62&	3.11&	2.95&	2.71&	-1.74&	3.04  \\            
% EchoTracker      & TAS-1K                      & 3.52&	3.37&	2.61&	2.69&	\textbf{-0.06}&	3.57&	3.85&	3.67&	3.09&	1.97&	\textbf{0.15}&	2.64  \\            
% MyoTracker       & TAS-1K                      & 1.59&	\textbf{0.61}&	1.91&	2.47&	-0.11&	3.25&	1.53&	\textbf{0.29}&	2.01&	1.49& -0.20 &	2.01  \\            
% TAS-Net          & TAS-1K                      & \textbf{1.52}&	1.02&	\textbf{1.64}&	\textbf{2.04}&	-0.28&	\textbf{2.71}&	\textbf{1.39}&	0.91&	\textbf{1.61}&	\textbf{1.42}&	-0.33&	\textbf{1.86}  \\            

% \midrule
% \midrule
%%%%%%% RESULTS WITH MYOTRACKER sequence length 64 (like in the paper)
\multicolumn{2}{c}{Reference}	& -		& -		& -		& 2.02	& -0.10	& 2.77	& -		& -		& -		& 1.78	& 0.78	& 2.15  \\
EchoTracker      & Real       	& 1.59	& 0.59	& 1.94	& 1.98	& -0.03	& 2.57	& 1.50	& 0.95	& 1.85	& 1.14	& 0.34	& 1.47  \\
\midrule
% EchoTracker      & Burman     	& 5.78	& 5.72	& 3.02	& 2.96	& 0.73	& 3.82	& 6.21	& 6.18	& 3.28	& 2.27	& 0.87	& 2.99  \\
% MyoTracker       & Burman     	& 3.30	& 2.99	& 2.57	& 2.37	& 0.20	& 3.10	& 4.53	& 4.45	& 2.69	& \textbf{1.40}	& 0.41	& 1.88  \\
% TAS-Net          & Burman     	& 2.60	& 1.36	& 3.00	& 2.54	& 0.30	& 3.34	& 3.62	& 3.11	& 2.95	& 2.71	& 1.74	& 3.04  \\
EchoTracker      & TAS-1K     	& 3.52	& 3.37	& 2.61	& 2.69	& \textbf{0.06}	& 3.57	& 3.85	& 3.67	& 3.09	& 1.97	& \textbf{-0.15}	& 2.64  \\
MyoTracker       & TAS-1K     	& 2.12	& 1.67	& 2.17	& 2.62	& 0.23	& 3.39	& 2.31	& 1.96	& 2.47	& 1.65	& 0.43	& 2.21  \\
TAS-Net          & TAS-1K     	& \textbf{1.52*}	& \textbf{1.02}	& \textbf{1.64}	& \textbf{2.04*}	& 0.28	& \textbf{2.71}	& \textbf{1.39*}	& \textbf{0.91}	& \textbf{1.61}	& \textbf{1.42*}	& 0.33	& \textbf{1.86}  \\

\bottomrule      
\end{tabular*}

\end{table*}

An ablation study using SSHF evaluates the individual contributions introduced in TAS-Net to improve upon the CoTracker framework for point tracking. Results are reported in \Cref{tab:ablation}. Performance was evaluated both in terms of geometric (MTE, folds) and clinical (GLS) metrics. Improved performance in terms of MTE is observed across all proposed contributions. Both the continuous visibility loss and the mesh embedding lead to improved GLS and MAE. Notably, the mesh embedding is the only contribution that substantially reduces mesh folding, highlighting the importance of enabling \mbox{TAS-Net} to explicitly distinguish points based on their mesh indices. In comparison, MyoTracker, trained on the same data, produces mesh folding in $63\pm6$\% of SSHF videos.

%We performed an ablation study evaluating the different contributions we proposed in TAS-Net to improve upon CoTracker method for echocardiographic video point tracking. We analyze the effect of the bidirectional sliding window, continuous visibility loss, and mesh embedding. Results are presented in \Cref{tab:ablation}. 

% We can observe improved performance on the MTE across all proposed contributions. The visibility loss and mesh embedding both contribute to improved GLS MAE. The mesh embedding is the only contribution that considerably reduces folds. This shows the impact of allowing TAS-Net to distinguish between points by the mesh index. In comparison, MyoTracker, trained on the same data, produced mesh folds in $63\pm6$\% and $70\pm5$\% of sequences for the SSHF and HUNT datasets, respectively. 

%%%%%%%%%%%%%%%%%%%%%%%%%%%%%%%%%%

%\subsection{State of the art}
\subsection{Comparisons with state-of-the-art methods}

TAS-Net was compared with Echotracker and MyoTracker. These methods were trained on both the simulated dataset proposed by Burman et al. and our proposed TAS-1K dataset.

\noindent\textbf{Motion estimation scores.} \Cref{tab:sota_geometric} displays the MTE scores obtained by each method on SSHF, HUNT and HUNT-A on the A2C, A3C and A4C cardiac views. For all evaluated methods, training on the \mbox{TAS-1K} synthetic dataset leads to the best performance across the validation datasets and views, indicating the effectiveness of the proposed simulation strategy. Also, \mbox{TAS-Net} outperformed \mbox{MyoTracker} and \mbox{EchoTracker}, regardless of the synthetic dataset used for training. These results support the effectiveness of the proposed architectural choices, in particular the use of a CoTracker-based architecture leveraging pre-trained weights on a large-scale dataset.

Finally, we compared 
%the performance of 
\mbox{TAS-Net} trained on the \mbox{TAS-1K} synthetic dataset with 
%that of 
EchoTracker trained on real echocardiographic videos, using myocardial motion references extracted from EchoPac. Although this comparison is inherently biased, \replaced{as both EchoTracker's training dataset and our validation dataset (two distinct datasets) use EchoPac-derived annotations}{as EchoTracker was trained using EchoPac-derived annotations that were also employed to generate the reference for the validation datasets}, 
the two methods show comparable performance. %we observe comparable performance between the two methods. 
This %result 
suggests that the proposed approach 
can be competitive %can achieve competitive performance 
when trained exclusively on synthetic data. \Cref{fig:worst_sample} presents a qualitative comparison of \mbox{TAS-Net} and \mbox{EchoTracker}, trained on real data, focusing on the worst-performing samples in terms of MTE. 
It %The figure 
shows that \mbox{EchoTracker} exhibits severe tracking errors, whereas \mbox{TAS-Net} remains closer to the EchoPac reference, with discrepancies that are more limited in magnitude.

\noindent\textbf{Global longitudinal strain scores.} \Cref{tab:gls} reports the performance of the different methods for peak GLS estimation on the test–retest datasets (combining SSHF and HUNT) and on the agreement dataset (HUNT-A). Consistent with the geometric metrics, training on the \mbox{TAS-1K} synthetic dataset yields the best results. Again, \mbox{TAS-Net} outperformed both \mbox{MyoTracker} and \mbox{EchoTracker}. Our model also produced comparable results compared to \mbox{EchoTracker} trained on real data with the reference motion extracted from EchoPac. Notably, the performance of \mbox{TAS-Net} is on par with the inter-observer variability measured in both the test–retest and agreement settings, suggesting good robustness under clinical acquisition conditions. This observation is further supported by the Bland–Altman plots shown in the first column of \Cref{fig:rs}, computed on the agreement dataset. These plots indicate a near-zero bias and limits of agreement below 5\% between GLS estimates obtained with \mbox{TAS-Net} and clinician references. %, which is close to the 3\% threshold commonly accepted in clinical practice. 

Both the MAE metric and Bland–Altman analysis show an increased variability between the test-retest setting and the agreement setting, showing the impact of the added variability induced by the probe position. However, in both cases TAS-Net exhibits a level of robustness to all forms of variability on par with the reference obtained with EchoPac.

% \Cref{tab:gls} displays the performance of the different methods in estimating the peak GLS on the test-retest datasets (combination of the SSHF and HUNT datasets) and the agreement dataset (HUNT-A). The same conclusions as the ones for the geometric scores can be made, namely the training on the \mbox{TAS-1K} synthetic dataset produced the best results and \mbox{TAS-Net} outperformed \mbox{MyoTracker} and \mbox{EchoTracker}. Remarkably, \mbox{TAS-Net} produced on pare with the inter-observer variability measured both in terms of test-retest and agreement. This illustrates the robustness and accuracy of our model in data acquired in clinical conditions.

% We also compare methods with respect to peak global longitudinal strain (GLS). In \Cref{tab:gls}, we present results with respect to the reference and in a test-retest setting on the combination of both the test-retest dataset and the HUNT agreement dataset. Once again, training on $\mathcal{D}^2_{S1 \shortto S2 \shortto S3}$ consistently improves clinical GLS accuracy compared with Burman et al. data. Furthermore, TAS-Net ($\mathcal{D}^2_{S1 \shortto S2 \shortto S3}$) provides the best overall GLS agreement among simulation-trained methods, and approaches the real-data EchoTracker baseline. MyoTracker performs simlilarly. 

\subsection{Regional Strain}

\begin{figure}[tp]
\centering
\includegraphics[width= \linewidth]{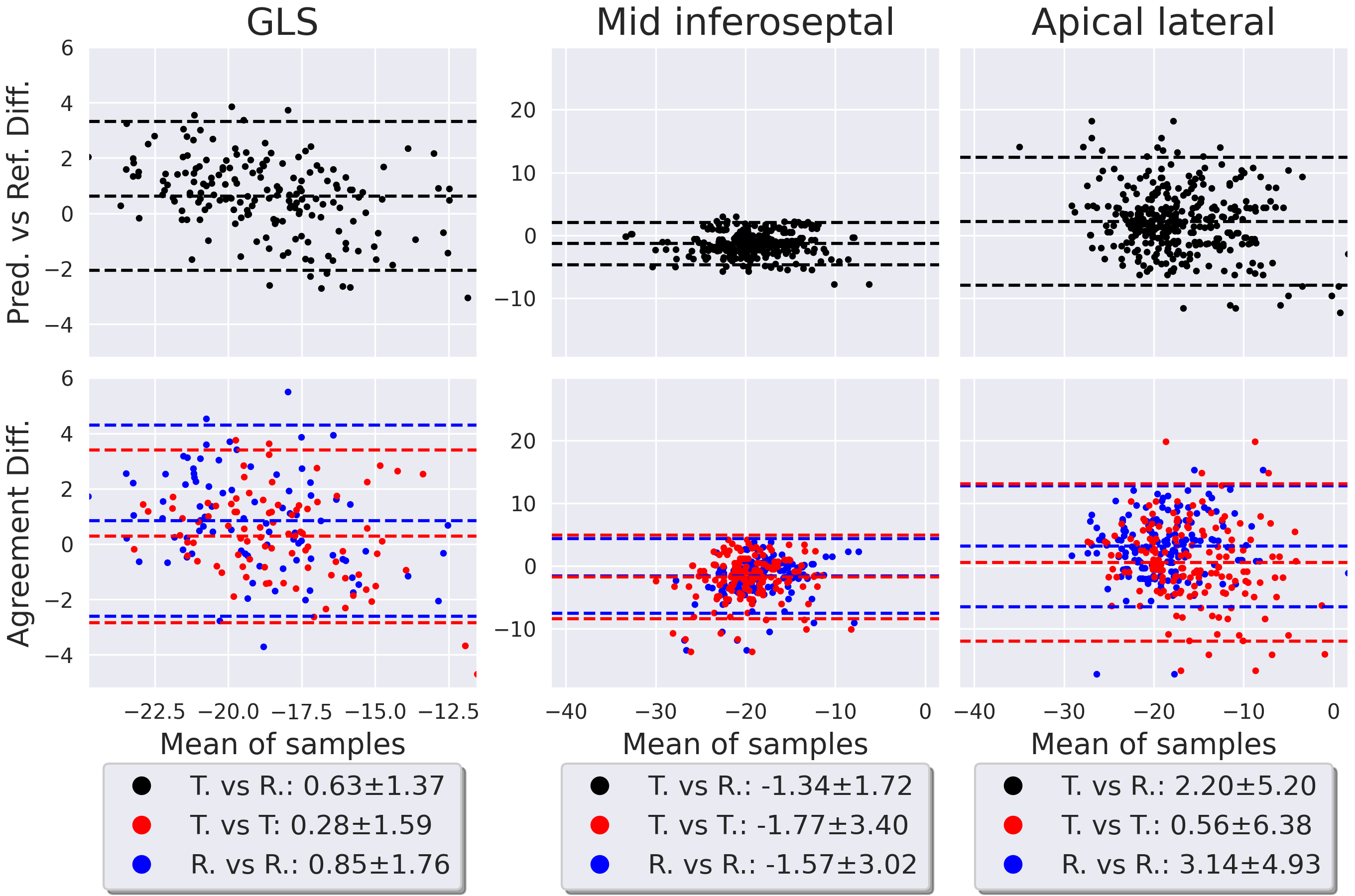}
\caption{Bland–Altman analysis of \mbox{TAS-Net} (T) for GLS and RLS (best and worst segments), evaluated against measurements from an expert using EchoPac (T vs. R) and in an inter-expert agreement evaluation setting (T vs. T and R vs. R) on the A4C sequences of the HUNT-A dataset.}
\label{fig:rs}
\end{figure}

We investigate the ability of \mbox{TAS-Net} to estimate regional longitudinal strain (RLS) from echocardiographic videos. 
The Bland–Altman plots given in~\Cref{fig:rs} compare %This analysis relies on Bland–Altman plots given in~\Cref{fig:rs} comparing 
RLS estimates obtained with the proposed model and with two expert clinicians on HUNT-A. 
\replaced{RLS was computed for the 18 segments (6 per view) following the AHA definition}{RLS was computed for the six myocardial segments from basal-septal to basal-lateral (RLS1 to RLS6) following the standard AHA definition.} %segmentation. 
We perform the analysis on HUNT-A, which measures the agreement of two experts using 
EchoPac %the EchoPac software 
on the same video acquisition. For each segment, three comparisons are considered: (i) \mbox{TAS-Net} 
vs. %versus
reference, where each reference is considered independently, and TAS-Net is initialized with the reference centerline at ES (T vs. R); (ii) a \mbox{TAS-Net} agreement measuring configuration, 
where %in which 
we compare the 
TAS-Net prediction %prediction of TAS-Net 
with the different initialization performed by each expert (T vs. T), and (iii) an expert agreement analysis, where two independent experts' references, obtained with  EchoPac, are compared (R vs. R).

% For each segment, three comparisons are considered: (i) \mbox{TAS-Net} versus Expert 1 using the same ES centerline initialization \textcolor{red}{(C vs R)}; (ii) a \mbox{TAS-Net} test–retest configuration, in which both the acquisition and ES centerline initialization were performed independently by two experts \textcolor{red}{(T vs T)}; and (iii) an expert test–retest analysis, where two independent experts performed their own acquisitions and RLS measurements using EchoPac \textcolor{red}{(R vs R)}.

\begin{figure}[tp]
    \centering
    \includegraphics[width=\linewidth]{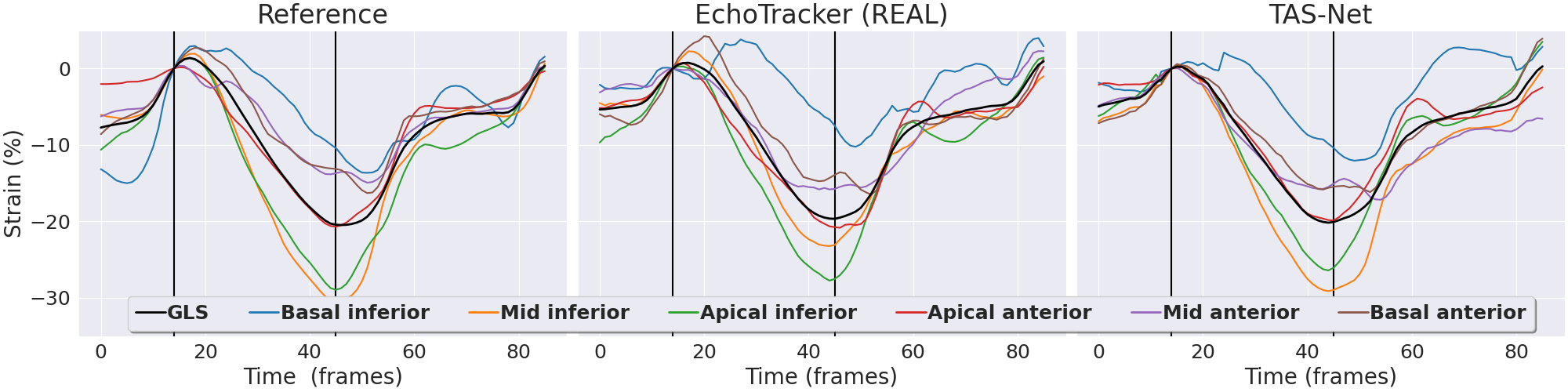}
    \caption{Regional strain 
    curves from %estimates by 
    the best-performing deep learning methods [middle and right]
    %, \mbox{TAS-Net} and EchoTracker trained on real data, 
    compared with those 
    from %obtained by 
    a clinical expert using EchoPac [left].}
    \label{fig:strain_map}
\end{figure}

An analysis of inter-expert variability across all three views reveals limits of agreement below 10\% for \replaced{mid-inferior segments}{RLS2}, 
$\sim$10\% % approximately 10\% 
for \replaced{basal and mid-anterior segments}{RLS1, RLS5, and RLS6}, and slightly above 10\% for \replaced{apical segments}{RLS3 and RLS4}, confirming the lower reproducibility of RLS %regional strain 
%measurements 
compared to GLS.
Across all segments, the variability observed between \mbox{TAS-Net} and Expert 1, as well as the test–retest variability of \mbox{TAS-Net}, is of the same order of magnitude as the inter-expert variability. \added{A qualitative example of RLS prediction is given in \Cref{fig:strain_map}.} These results indicate that the proposed approach can achieve reproducible performance comparable to that of expert clinicians, despite being trained exclusively on synthetic data.

% \begin{figure}[tp]
%     \centering
%     \includegraphics[width=\linewidth]{figures/34_2CH.png}
%     \caption{Regional strain estimates and strain maps produced by the best-performing deep learning methods, \mbox{TAS-Net} and EchoTracker trained on real data, compared with those obtained by a clinical expert using EchoPac.}
%     \label{fig:strain_map}
% \end{figure}

\section{Discussions and conclusions}

%\paragraph*{\textbf{Feasibility of synthetic-only training} }
\noindent\textbf{Feasibility of synthetic-only training.}
In this study, we show that the proposed strategies to enhance the realism of synthetic datasets, in terms of both speckle decorrelation and myocardial motion, enable 
training %the training of 
deep learning models with improved performance compared to existing synthetic datasets. In particular, the proposed multi-stage refinement strategy, incorporating progressively increasing levels of speckle realism, consistently outperformed the other evaluated configurations.
We also %Moreover, we 
demonstrate that designing a dedicated point-tracking model based on the CoTracker architecture, leveraging large-scale pre-trained weights and adapting it to the characteristics of echocardiographic videos, is 
key to achieve %a key factor in achieving 
highly competitive performance. Comparisons %with expert variability 
on the test–retest and agreement datasets show that \mbox{TAS-Net} achieves performance comparable to inter-expert variability for both motion 
%estimation 
and peak GLS estimation. To the best of our knowledge, such a level of agreement between a synthetic-only trained model and expert variability has not been previously reported. 
Our approach therefore %These findings suggest that the proposed approach 
has 
strong %the 
potential to improve the robustness of peak GLS measurements under clinical acquisition conditions.

\noindent\textbf{Generalization across dataset, views and pathologies.}
A key contribution of this work is to demonstrate that the generation of realistic synthetic datasets enables the generalization of deep learning models to real echocardiographic videos. 
This generalization ability was further evaluated across views. While \mbox{TAS-1K} only includes synthetic videos from the A4C and A2C views, the validation datasets comprise all three standard 
%clinical 
views 
(A4C, A2C, A3C).
%, namely A4C, A2C, and A3C.
As shown in \Cref{tab:sota_geometric}, \mbox{TAS-Net} achieves comparable performance across views, indicating consistent behavior despite the absence of A3C data during training. Also, \mbox{TAS-Net} attains performance comparable to \mbox{EchoTracker}, which is trained on real data incorporating all three views. This underlines that our approach generalizes well across both datasets and echocardiographic views and diseases. Finally, TAS-Net’s GLS error with respect to the EchoPac reference varies by at most 0.6\% across all disease categories in the evaluation dataset (hyperlipidemia, hypertension, diabetes, angina, and coronary artery disease), suggesting good generalization across different clinical conditions.

\noindent\textbf{Deep learning methods for regional strain estimation.}
The inter-expert variability shown in~\Cref{fig:rs} highlights the limited reproducibility of regional strain %measurements 
across experts. While 
some segments (e.g. mid-inferior) %certain segments, such as mid-inferior segments, 
exhibit variability comparable to that observed for GLS, most segments present limits of agreement around 10\%, which remains high for routine clinical use.
\mbox{TAS-Net} achieves a level of accuracy and reproducibility comparable to clinical inter-expert variability. However, these results are influenced by the use of manual expert annotations at ES to initialize the tracking procedure. This initialization step inherently introduces human variability, both in the selection of the ES frame and in the delineation of the myocardial contour. As a result, the reported performance of \mbox{TAS-Net} incorporates this source of variability.
This limitation could be mitigated by replacing manual initialization with an automatic segmentation approach for both ES frame selection and initial contour placement. 
This %Although such an 
extension is beyond the scope of this study,  but  it will be investigated in future work.

\noindent\textbf{Limitations and future work.}
%The use of 
%EchoPac %EchoPac-derived 
%outputs as reference 
%is a limitation, %constitutes a limitation of this study, 
%as they 
EchoPac outputs 
do not provide a direct measurement of true myocardial motion. This %limitation 
is partially mitigated by restricting the evaluation to myocardial segments validated by clinicians and by assessing performance on both test–retest and agreement datasets.
Nevertheless, future work should include validation against independent reference modalities, such as cardiac MR or tagged MR imaging, to enable a more objective assessment of strain accuracy. %In addition, we plan to investigate and evaluate the accuracy and reproducibility of a fully automated end-to-end framework by integrating automatic segmentation.

\bibliographystyle{IEEEtran}
\bibliography{ref}

\end{document}